\documentclass[preprint,3p,12pt]{elsarticle}

\usepackage{amssymb}
\usepackage{amsmath}
\usepackage{booktabs}
\usepackage{hyperref}
\usepackage{algorithm}
\usepackage{algpseudocode}
\usepackage{longtable}

\usepackage{amsthm}
\usepackage{xurl} 

\usepackage{multirow}
\usepackage{subcaption} 
\usepackage{adjustbox}
\usepackage{mathtools}
\usepackage{color}
\usepackage[normalem]{ulem}

\usepackage{amsmath}
\newcommand{\mathbbm}[1]{\text{\usefont{U}{bbm}{m}{n}#1}} 

\usepackage{verbatim} 

\journal{Expert Systems with Applications}

\begin{document}

\begin{frontmatter}


\author[dtu]{Bastian Schmidt Jørgensen\corref{cor1}}
\cortext[cor1]{Corresponding author. \textit{Email address:} {bassc@dtu.dk}.
\textit{Phone:} {+45 45 25 25 25}
}
\fntext[jan2]{\textit{Email address:} {jkmo@dtu.dk}.}
\fntext[peter2]{\textit{Email address:} {pnys@dtu.dk}.}
\fntext[henrik2]{\textit{Email address:} {hmad@dtu.dk}.}

\title{Sequential Methods for Error Correction of Probabilistic Wind Power Forecasts} 


\affiliation[dtu]{organization={DTU Compute},
            addressline={Bygning 303B}, 
            city={Kgs. Lyngby},
            postcode={2800}, 
            country={Denmark}}
            
\affiliation[peter]{organization={QUENT ApS},
             addressline={Svanemøllevej 41},
             city={Hellerup},
             postcode={2900},
             country={Denmark}}

\author[dtu,jan2]{Jan Kloppenborg Møller} 
\author[dtu,peter,peter2]{Peter Nystrup}
\author[dtu,henrik2]{Henrik Madsen}

\begin{abstract}
Reliable probabilistic production forecasts are required to better manage the uncertainty that the rapid build-out of wind power capacity adds to future energy systems. In this article, we consider sequential methods to correct errors in wind power production forecast ensembles derived from numerical weather predictions. 
We propose combining neural networks with time-adaptive quantile regression to enhance the accuracy of wind power forecasts. We refer to this approach as Neural Adaptive Basis for (time-adaptive) Quantile Regression or NABQR. 
First, we use NABQR to correct power production ensembles with neural networks.
We find that Long Short-Term Memory networks are the most effective architecture for this purpose.
Second, we apply time-adaptive quantile regression to the corrected ensembles to obtain optimal median predictions along with quantiles of the forecast distribution. 
With the suggested method, we beat state-of-the-art methods and achieve accuracy improvements up to 40\% in mean absolute terms in an application to day-ahead forecasting of on- and offshore wind power production in Denmark. In addition, we explore the value of our method for applications in energy trading.
We have implemented the NABQR method as an open-source Python package to support applications in renewable energy forecasting and future research.
\end{abstract}


\begin{graphicalabstract}
\includegraphics[width=1\textwidth]{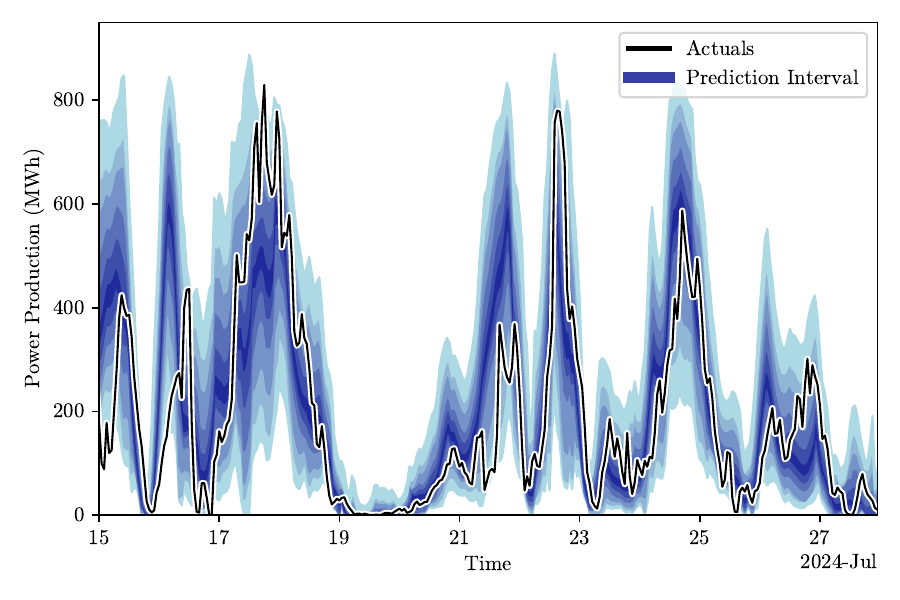}
\end{graphicalabstract}

\begin{highlights}
\item Developed neural networks for data-driven updates of wind power ensembles
\item Applied time-adaptive quantile regression on corrected ensembles
\item Improved wind power production forecasts by up to 40\% in a mean absolute sense
\item Published well-documented open-source Python package: \texttt{nabqr}
\end{highlights}


\end{frontmatter}




\section*{Abbreviation List}
See Table \ref{tab:acronyms}. 
\begin{table*}[!ht]
    \centering
    \begin{tabular}{ll}
        \textbf{Abbreviation} & \textbf{Definition} \\
        \hline
        QS & Quantile loss function Score \\
        HPE & High Probability Ensemble (or control member) \\
        QRF & Quantile Random Forest or Quantile Regression Forest \\
        QGB & Quantile Gradient Boosting \\
        NWP & Numerical Weather Prediction \\
        CRPS & Continuous Ranked Probability Score \\
        LSTM & Long Short-Term Memory (network) \\
        TAQR & Time-Adaptive Quantile Regression \\
        QRNN & Quantile Regression Neural Network \\
        NABQR & Neural Adaptive Basis for (time-adaptive) Quantile Regression \\
        ECMWF & European Center for Medium-Range Weather Forecast \\
    \end{tabular}
    \caption{List of abbreviations and their definitions.}
    \label{tab:acronyms}
\end{table*}

\newpage
\section{Introduction}
\label{intro}
Improving the accuracy of renewable energy production forecasts is crucial to maintaining the stability and reliability of the power grid. In 2023, 82\% of the electricity produced in Denmark came from renewable energy sources, with wind accounting for 54\% \citep{energistyrelsen2024}. The production from renewable energy sources varies greatly depending on weather conditions. Accurate forecasts allow grid and energy system operators to optimize the use of existing renewable energy resources and minimize costs and emissions associated with running additional fossil fuel-based (backup) power plants. Accurate and reliable forecasts are also needed to activate demand-side flexibility, balance supply and demand, and provide grid services \citep{lessfossil0,madsen2025a}.

The probabilistic forecast distribution is of great interest to grid operators fighting to prevent blackouts and to asset owners and energy traders looking to minimize volatility, do optimal bidding given asymmetric cost functions, or accurately estimate Conditional Value at Risk (CVaR), which quantifies potential extreme losses in the left tail of a return distribution. \citep{dahlgren2003risk,costfunc_maciejowska2024}. Renewable energy producers are exposed to significant financial losses during volatile price spikes in the imbalance market. Better probabilistic forecasts improve CVaR estimates, helping producers reduce financial risks during volatile periods \citep{cVar2023}.

In practice, a common approach to estimate the forecast density and quantify the uncertainty is to convert the 51 weather ensembles from the European Centre for Medium-Range Weather Forecasts (ECMWF) into power production ensembles. The ECMWF calculates these 51 ensembles by slightly perturbing the initial conditions on their control ensemble and projects these ensembles forward in time by advanced PDE systems \citep{ecmwf_website}. The underlying and implicit assumption is that with precise initial conditions, one could predict tomorrow’s weather exactly, as PDE systems are deterministic and their solutions are deterministic functions of time and space. However, since we do not know the initial weather conditions at all locations and since the system is continuously affected by stochastic disturbances, this inherently calls for error corrections to provide reliable probabilistic forecasts. 

Quantile regression methods are often used with ECMWF's weather ensembles and power production ensembles as input to provide probabilistic forecasts. \citep{qr_SOTA} provided a state-of-the-art review of current quantile regression (QR) methods. They found that QR Deep Learning (QRDL) methods had the highest accuracy, but also the highest complexity level. Examples of QRDL methods include QR-LSTM, QR-deep neural network (QR-DNN), and QR-Convolutional neural network (QR-CNN). We employ a QR-LSTM network followed by two dense layers. Furthermore, we compare our final method to Quantile Regression Forest (QRF), which is a method with lower complexity and accuracy, and Quantile Gradient Boosting (QGB), which improved state-of-the-art methods in \citep{qgbsolar}. 

To improve the reliability of the forecast distribution and correct the errors that arise in weather ensembles and therefore also power production ensembles, \citep{taqrpaper} used time-adaptive quantile regression (TAQR) to model a number of quantiles. They found that TAQR was superior to QR for wind power production data. 

In \citep{windens}, the authors went from wind forecasts to wind ensembles and then quantile forecasts. Their approach included statistical modeling, as they argued that they could not go directly from the wind power ensembles to quantile forecasts. In other words, they hint that additional steps of modeling were needed to go from wind power ensembles to the final quantile forecasts. \citep{qrExtend} used quantile regression to extend their existing wind power production forecasts to probabilistic forecasts. They argued that analysis of the forecast error can be used to build a model for the quantiles. \citep{lauretProbsolar} found that using Numerical Weather Predictions (NWP) improved their quantile regression models. In this article, to keep the method general, our only input is power production forecast ensembles, thus the method works without other external inputs, such as NWP.

In \citep{nbeats}, N-BEATS was introduced as the first interpretable deep-learning time-series forecasting tool to beat traditional statistical methods. N-BEATS uses neural networks to learn a basis for the forecasting coefficients. In \citep{Cannon2011}, quantile regression was implemented with neural networks in the \verb|R| package called \verb|qrnn|. Neural networks designed to minimize the quantile loss between their predictions and the target values are commonly referred to as Quantile Regression Neural Networks (QRNNs) \citep{qr_SOTA}. In \citep{Bremnes2020}, the quantile functions were specified by linear combinations of Bernstein basis polynomials, while their coefficients were assumed to be related to ensemble forecasts utilizing a highly adaptable neural network. 

These articles, and the neural architecture they introduce, motivated us to suggest using neural networks in combination with a multidimensional quantile loss function. This leads to corrected ensembles, which can then serve as a basis for TAQR. In this way, we use neural networks to form the basis for a regression task, similar to what \citep{nbeats} did.


Our objective is to correct errors linked to the ECMWF's meteorological ensembles and provide improved probabilistic forecasts of wind power production in Denmark. To achieve this efficiently, we select TAQR, which requires a basis. Rather than relying on raw power production forecast ensembles, we correct them using neural networks, creating a calibrated data-driven basis for TAQR. Thus, we propose to combine neural networks with TAQR to correct wind power production ensembles and turn them into probabilistic wind power forecasts through a sequence of methods. We refer to this approach as
Neural Adaptive Basis for (time-adaptive) Quantile Regression or NABQR. To the best of our knowledge, using neural networks to update the basis matrix for a regression task, specifically the TAQR algorithm, is a novel approach.

A common problem in quantile regression is the crossing of quantiles. Our method effectively eliminates quantile crossings during ensemble corrections and provides a robust basis for the TAQR model that predicts tomorrow’s power production quantiles. For the implementation of the TAQR model, we leverage the simplex method for quantile regression, as described in \citep{taqrpaper}. This improves computational efficiency, since we often only need a few iterations in the simplex method to converge. Our approach is distinguished by its integration of neural networks to form a suitable data-driven basis. 

We developed a Python package that incorporates the suggested neural network-based correction methodology and the TAQR algorithm. The code, openly shared on GitHub\footnote{\url{https://github.com/bast0320/nabqr}}, facilitates reproducibility and enables further research in this domain.

The remainder of the article is organized as follows. Section~\ref{sec:data} presents the data and describes the data-cleaning approach used. In Section \ref{sec:method}, the methods are presented. The results are presented in Section~\ref{sec:results} followed by an application to energy trading in Section~\ref{sec:applications}. We discuss the results and propose ideas for future work in Section~\ref{sec:discussion}. Finally, we conclude the article in Section~\ref{sec:conclusion}.

\section{Data}
\label{sec:data}
To properly motivate the methods, we present the data here. We use power production in DK1 and DK2 from off- and onshore wind farms. Denmark is split into two price zones. DK1 is West Denmark and DK2 is East Denmark. 

The actual observations of power production are available from the Danish transmission system operator Energinet as 5-minute observations \citep{energidataservice_website}. We aggregate these observations to hourly resolution to match the frequency of the forecast ensembles. The 5-minute observations are reported in megawatts (MW). However, since we are focusing on hourly observations, the units of MW and megawatt-hours (MWh) can be used interchangeably. Each hourly data point represents the average production over one hour.

\begin{figure}
    \centering
    \includegraphics[width=1\linewidth]{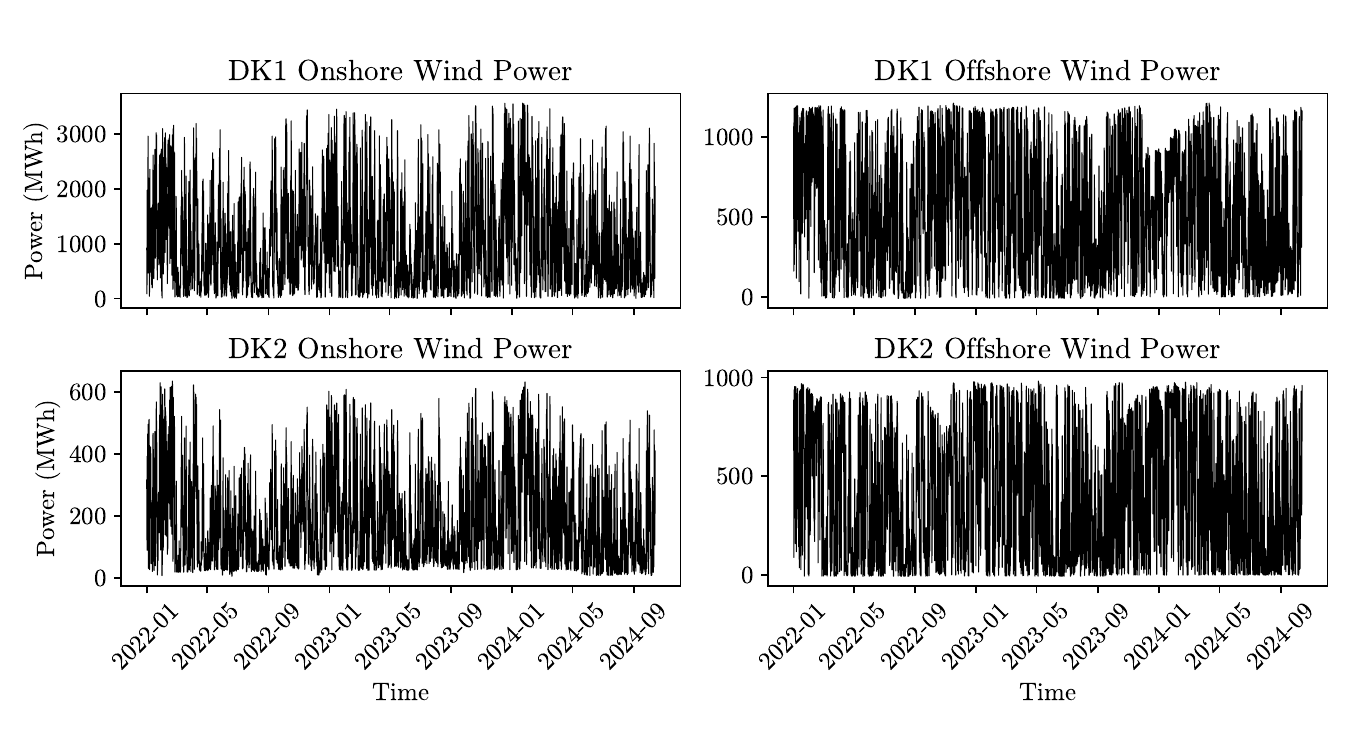}
    \caption{Hourly observations for DK1 and DK2 off- and onshore wind power production in MWh.}
    \label{fig:rawdata}
\end{figure}

The data set spans the period 2022-01-01 to 2024-10-12 with a total of 24381 hourly observations. From the aggregated observations shown in Fig.~\ref{fig:rawdata}, we see less onshore wind power production during the period from May to August, where the average wind speed is lower than in winter and autumn in Denmark \citep{ANDRESEN2015}. Ultimo 2023 the installed wind power production capacity in Denmark for off- and onshore was approximately 6899 MW (corresponding to 60 TWh produced if fully utilized year round), while 19.5 TWh was produced in 2023 \citep{windenergistyrelsen}. 

\begin{figure}
    \centering
    \includegraphics[width=1\linewidth]{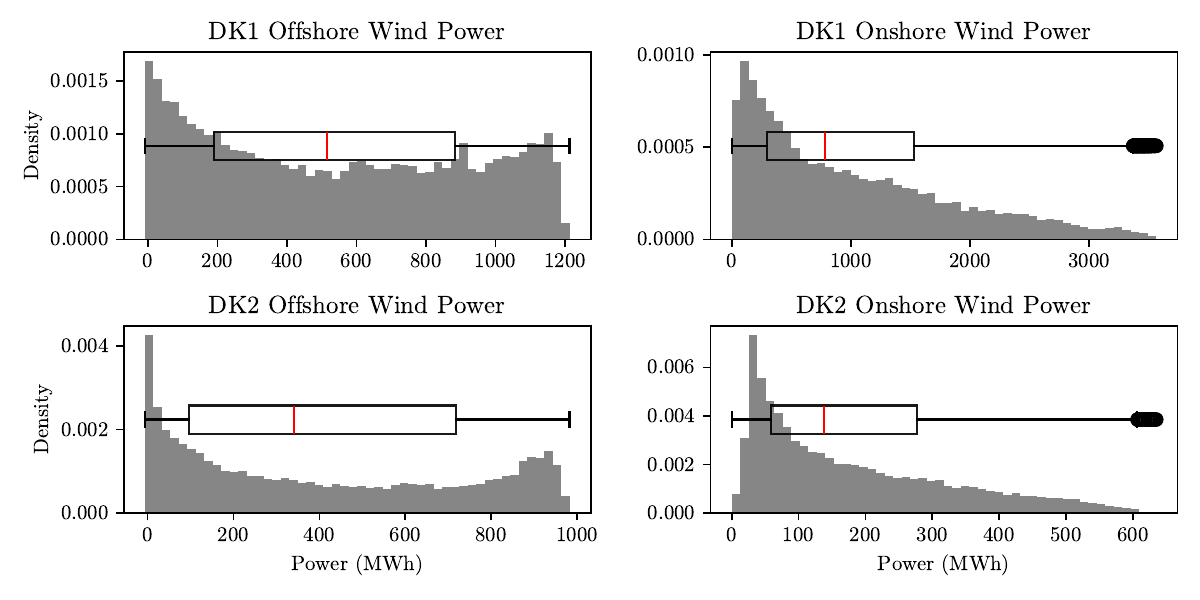}
    \caption{Density plot with 50 bins of hourly observations with overlayed boxplot for DK1 and DK2 off- and onshore wind power production.}
    \label{fig:densityplot}
\end{figure}

In Fig.~\ref{fig:densityplot}, the densities of the actual observations are plotted alongside their boxplot. Offshore wind power production has a U-shaped density curve, indicating that offshore wind turbines are often running at close to max or zero power. Onshore shows a right-skewed distribution indicating many observations with power output in the lower range and only very few observations at the maximum values. This can make it more difficult to estimate the higher quantiles for onshore wind power production with quantile regression.

Forecasted {weather ensembles} are available from the ECMWF \citep{ecmwf_website}. ECMWF generates 50 distinct scenarios of potential weather outcomes by perturbing initial conditions slightly on their {High Probability Ensemble}, called the {Ensemble Forecast Suite} (ENS) \citep{ens}. Thus, we have in total 51 ensembles. 
When averaged over many forecasts each of the perturbed ensembles has lower skill than the unperturbed control ensemble, but any individual forecast might show a higher skill \citep{ens}.

The weather ensembles are transformed into {forecast ensembles} for power production through state-of-the-art model(s) by a commercial wind power forecast provider. Due to the proprietary nature of the forecasting business, we have no knowledge of the specific model(s) used.

The forecast ensembles represent 51 scenarios for wind power production in DK1 and DK2. The forecast ensembles are referred to as the \textit{ensembles}. The ensembles for day $d$ are available before 12:00 CET at day $d-1$, thus the forecast horizon is 12 to 36 hours.

\begin{figure}
    \centering
    \includegraphics[width=1\linewidth]{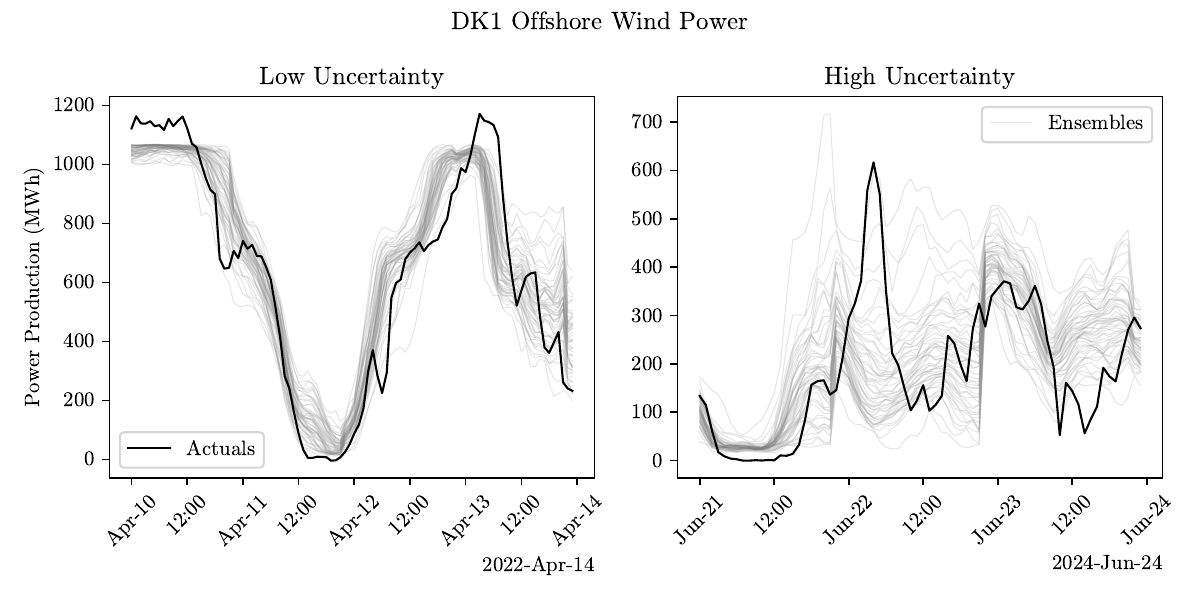}
    \caption{Wind power production for DK1 offshore for two periods with low uncertainty (left) and high uncertainty (right). Note that the ensembles have been sorted for this figure.}
    \label{fig:uncertaintyensembles}
\end{figure}

In Fig.~\ref{fig:uncertaintyensembles}, two different periods with low and high variability in the ensembles are shown for DK1 offshore wind power. Low variability in the ensembles implies high confidence in the forecast. However, these narrow ensembles fail to fully cover the observed wind power production, indicating an underestimation of variability. Furthermore, they do not capture the maximum output. In the right figure, a larger spread appears appropriate given the increased variability and less predictable nature of the actual wind power production during this time. The observations, while exhibiting rapid changes, generally remain within the ensemble bounds, but the timing of the first peak is too early in the ensembles compared to the actual observations.

\subsection{Data cleaning}
Agreements between Denmark and Germany have allowed countertrading between West Denmark (DK1) and Northern Germany. The result is that sometimes offshore wind turbines in primarily DK1 have been curtailed. Going forward the curtailment issue will be less pronounced and is already fading due to new installations and expansions in the German energy network since July 2023 \citep{energinet_newdeal}. Since the power production ensembles cannot capture these sudden drops, the TAQR algorithm would have no way of predicting this, since it is purely data-driven. Consequently, we want to exclude these periods both from training and evaluation of our method. Incorporating additional inputs into the TAQR algorithm—such as countertrade volumes—could improve curtailment modeling. However, because countertrade volumes are not available on a day-ahead basis, they cannot be integrated into our forecasting methodology.

We exclude observations using a filter under the following conditions: If the countertrade volume exceeds 1700 MW before and after a period where it drops below 25 MW, or if the spot price is negative. To account for transitions between these periods, we extend the filter by padding each period flagged for removal by two zeros to each side, corresponding to a total of four hours of padding. 
The spot price refers to the price of electricity in the day-ahead auction. When the spot price is negative, producers shut down production to avoid paying to produce electricity. Danish wind farm producers attempt countertrading with Germany when production exceeds the capacity of the Danish energy grid. If further countertrading is not possible, curtailment becomes the last resort. Germany pays high prices to curtail Danish wind turbines because the bottleneck lies within their grid \citep{energinet_newdeal}.

The threshold of 1700 MW is chosen because curtailment is likely when the countertrade volume exceeds this level before dropping near zero and then rising above 1700 MW again. The 25 MW threshold is used as a small buffer above zero since countertrade volume is not always exactly zero during these periods of countertrading. 

\begin{figure}
    \centering
    \includegraphics[width=1\linewidth]{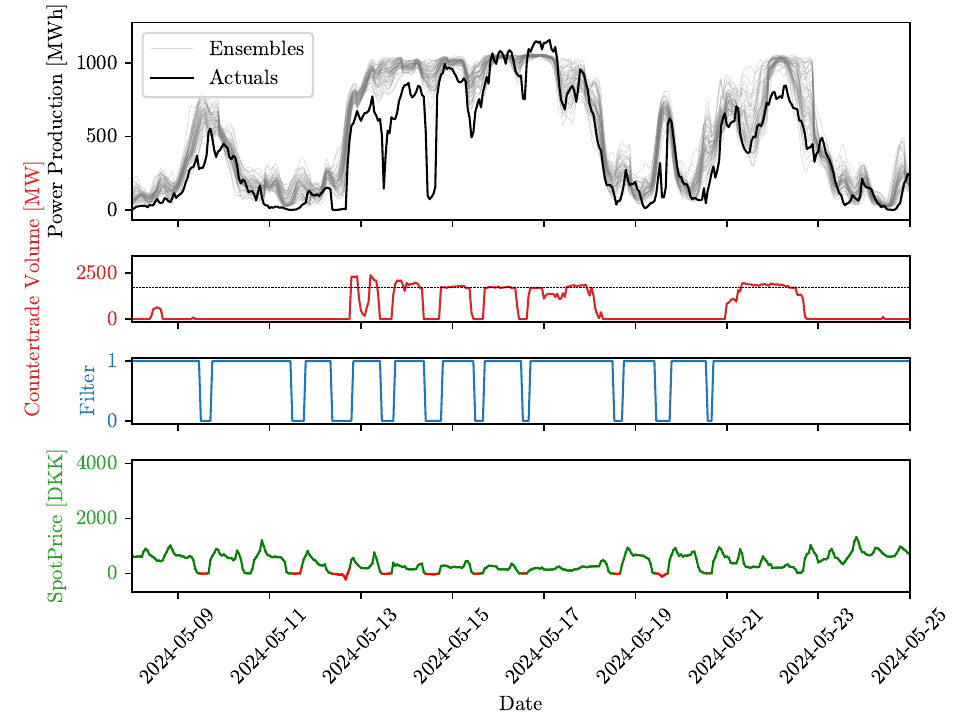}
    \caption{Spot price, countertrade volume, actuals observation, and ensembles shown alongside the filter. We remove observations where the filter is zero.}
    \label{fig:dk1curtail}
\end{figure}

Fig.~\ref{fig:dk1curtail} shows the (unfiltered) actual observations overlaid on the ensembles as well as the countertrade volume, spot price, and the filter. Observations are removed wherever the filter value is zero. Notably, sharp drops in wind power production are observed during periods of high countertrading and negative spot prices, which is what the filter corrects. The filter is only based on spot prices in the period 01-01-2022 to 01-07-2023 since data on special regulation was not obtainable in that period.

\begin{figure}
    \centering
    \includegraphics[width=1\linewidth]{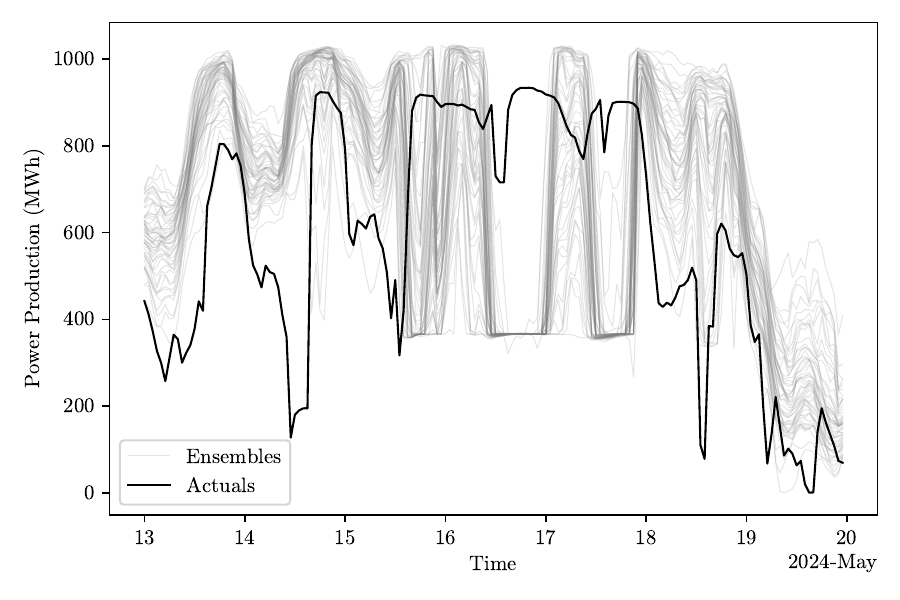}
    \caption{Actual observations and original ensembles for DK2 offshore wind power production. We remove observations where the ensembles group together around 365 MWh.}
    \label{fig:ensembleproblem}
\end{figure}

In addition to curtailment data errors also occur. Fig.~\ref{fig:ensembleproblem} shows a data error for DK2 offshore ensembles. The ensembles group together around 365 MWh for no apparent reason. We do not want our model to mistake data errors for actual data, so we remove observations in the entire dataset if more than 9 ensembles at time $t$ is less than 370 and greater than 358. As with the first filter, we extend the filter to capture the entire data error by a padding of two zeroes to each side. This data error was found by visual inspection.

\subsection{Training and Test Set}
\label{sec:set}
The data is divided into a training and test set. The training set is further divided into a training set for neural network training, the "warm start" \citep{Nielsen1999} of parameters for the TAQR algorithm and the initial observations in the TAQR algorithm. Table \ref{tab:traintestset} shows the start and end date of these periods as well as the amount of hourly observations they constitute. 

\begin{table*}[]
\centering
\begin{tabular}{@{}lllll@{}}
\toprule
                              &                             & Start period & End period & Hours \\ \midrule
\multicolumn{1}{l|}{\multirow{3}{*}{Training Set}} & Neural network & 2022-01-01 & 2023-08-09 & 14040 \\
\multicolumn{1}{l|}{}         & Initial parameters for TAQR & 2023-08-10   & 2023-08-18 & 192 \\
\multicolumn{1}{l|}{}         & Initial elements for TAQR   & 2023-08-19   & 2024-03-12  & 4944\\
\hline
\multicolumn{1}{l|}{Test Set} & Combined                    & 2024-03-13   & 2024-10-12  & 5112\\ \bottomrule
\end{tabular}
\caption{Overview of training and test sets.}
\label{tab:traintestset}
\end{table*}

\section{Methodology} \label{sec:method}

\subsection{Overview of NABQR}

\begin{figure}[h]
    \centering
    \includegraphics[width=\linewidth]{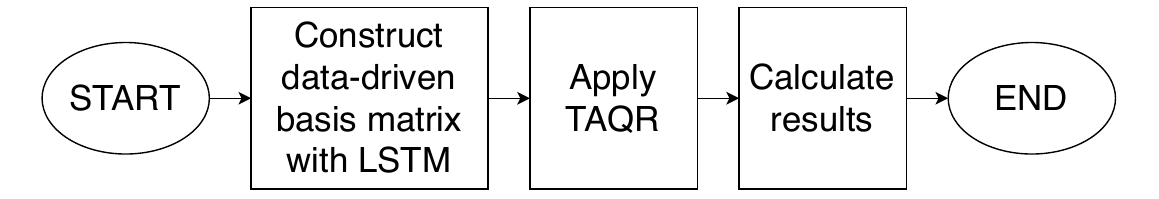}
    \caption{Brief flowchart to illustrate NABQR.}
    \label{fig:flowchart_simple}
\end{figure}

In this section, we present an overview and motivation of the steps in our Neural Adaptive Basis for Quantile Regression (NABQR) method. Fig.~\ref{fig:flowchart_simple} presents a flowchart of the overall steps in NABQR.

Quantiles of the forecast distribution are of interest. Quantile regression has the advantage that it can model an arbitrary number of quantiles. Instead of running a new complete quantile regression every day, to reduce the number of FLOPS and increase the speed, we will use an adaptive approach. We apply Time-Adaptive Quantile Regression (TAQR) \citep{taqrpaper}.

The TAQR algorithm uses the simplex method. This method is fast and relies on the previous time step solution as start guess for the next iteration. This is beneficial to us, since the distribution of our 5000-hour sliding look-back window does not change much between time steps. The simplex method needs a design matrix, $\mathbf{X}$, to perform the regression.

We have created an updated version of \citep{taqrreport} in Python, and the code is available on GitHub and can be installed by \texttt{pip install nabqr}.

Instead of using the ensembles directly in the design matrix, we perform a data-driven correction of the ensembles in $\mathbf{X}$. We will use neural networks to correct the ensembles, as outlined in section \ref{sec:nn}. To the best of our knowledge, neural networks have not been used like this before in the literature. By training our neural networks to best model the specified quantiles, we obtain an optimal – given the specific neural architecture – data-driven basis for the TAQR algorithm. 

\subsection{Neural networks}
\label{sec:nn}
Neural networks are chosen since they allow for highly flexible relations between covariates, support for non-linearities and allow estimation of all quantiles simultaneously. We minimize the quantile loss between the targets and the outputs, as normally done in QRNNs \citep{qr_SOTA}. We have experimented with different neural networks ranging from simple Feed Forward Neural Networks to Long Short-Term Memory (LSTM) networks to more advanced Transformers. We got the best results with LSTM networks, which we employed to update the design matrix.

Transformers leverage self-attention to capture complex temporal patterns; however, this requires substantial amounts of training data  \citep{finxter2023transformerlstm}. In the Transformer networks we applied in this article, the trainable parameter count ranged from 3 to 11 million, which we did not have enough training data for. The reason for the LSTM network being the best in this case is twofold. First, LSTMs are efficient for smaller datasets, effectively modeling shorter sequences with reduced memory needs compared to Transformers. Second, we observed that the LSTMs generalize better, and we know Transformers are prone to overfitting due to the high parameter count \citep{scispace2023cases}.

\subsubsection{LSTM networks}
LSTM networks have emerged as a powerful variant of recurrent neural networks (RNNs), designed specifically to mitigate the challenges posed by vanishing and exploding gradients, which hinder the capture of long-term dependencies in sequential data \citep{lstm}. LSTM networks are distinguished by their unique architecture, which incorporates gating mechanisms to selectively retain or discard information at each time step, making them particularly well-suited for time series prediction.

An LSTM unit contains three core gates: the input, forget, and output gate. These gates regulate the flow of information into and out of the memory cell, ensuring that relevant information is retained over time while irrelevant or outdated information is discarded. At each time step, the cell state — representing the long-term memory — is updated through a combination of input modulation and the forget mechanism, while the hidden state — reflecting the short-term memory — acts as the output of the LSTM module.

Mathematically, the gates and cell state update are defined as follows.

   \textit{The input gate} determines how much of the new input should be integrated into the cell state. This is governed by
   \[
   i_t = \sigma(W_i [h_{t-1}, x_t] + b_i),
   \]
   where \(i_t\) is the input gate activation, \(h_{t-1}\) is the previous hidden state, \(x_t\) is the current input, and \(\sigma\) denotes the sigmoid function. Generally, $W_{[\cdot ]}$ and $b_{[\cdot]}$ denotes the weight matrix and bias vector for gate $[\cdot]$, respectively. For dimension specifications, see Table \ref{tab:dimensionLSTM}.

\textit{The forget gate} decides what portion of the previous cell state should be retained. It is defined as
   \[
   f_t = \sigma(W_f [h_{t-1}, x_t] + b_f),
   \]
where \(f_t\) controls the retention of the previous cell state \citep{forgetgate}.

\textit{The cell state} is updated by combining the retained portion of the previous cell state with the new candidate cell state, modulated by the input gate
   \begin{align*}
   \tilde{C}_t = &\tanh(W_c [h_{t-1}, x_t] + b_c),\\
   C_t = & f_t \odot C_{t-1} + i_t \odot \tilde{C}_t,
   \end{align*}
where \(C_t\) is the updated cell state, \(\odot\) denotes the element-wise product, and \(\tilde{C}_t\) is the candidate cell state.

\textit{The output gate} regulates how much of the current cell state is output as the hidden state for the next time step
\begin{align*}
   o_t &= \sigma(W_o [h_{t-1}, x_t] + b_o), \\
   h_t &= o_t \odot \tanh(C_t),
\end{align*}
where $h_t$ is the current hidden state, which serves as both the output of the LSTM unit and the input to the next unit \citep{olah2015understanding}.

In the end, $h_t$ is sent through the MultiLayer Perceptron (MLP), which includes two hidden layers with 20 neurons each: Dense 1 and Dense 2 in Table \ref{tab:lstm_model_selection}. The Keras documentation refers to a model of this type as a \textit{Sequential Model} \citep{tensorflow}.

This architecture allows LSTM networks to excel in capturing long-term dependencies, making them particularly effective for a variety of applications that involve sequential data. The gating mechanisms enable LSTMs to selectively remember or forget information as needed, significantly improving their performance compared to traditional RNNs in tasks such as time series forecasting. This is excellent for our purpose of having a data-driven adaptive basis.

In practice, the LSTM network's parameters are optimized using backpropagation through time (BPTT), where the gradient of the loss function is propagated backward across time steps, updating the weights associated with each gate, \citep{bppt1,bppt2,bppt3}.
We build the FFNN and LSTM networks in TensorFlow \citep{tensorflow} and the Transformer in PyTorch \citep{pytorch}.

\textit{Loss function for LSTM network} \\
Assume $y \in \mathbb{R}^N$ are the true observations and $\hat{y} \in \mathbb{R}^N$ are our predictions for time $1,2,...,N$. Then the single quantile loss for a given quantile $q$ is
\begin{align*}
L_q(y_t, \hat{y}_t) = \max(&q \cdot (y_t - \hat{y}_t), (1-q) \cdot (y_t - \hat{y}_t)),
\end{align*}

Then the combined quantile loss for a set of quantiles $\{q_i\}_{i=1}^Q$, where $Q$ is the number of quantiles, to time $t$, is
\begin{align*}
L(y_t, \hat{y}_t) &= \frac{1}{Q}\sum_{i=1}^Q L_{q_i}(y_t^{(q_i)},\hat{y}_{t}^{(q_i)}),
\end{align*}
where $y_t^{(q_i)}$ is the $q_i$'th quantile of $[\mathbf{X}_t^T,y_t]^T$ 
and $\hat{y}_{t}^{(q_i)}$ is the predicted value for the $q_i$'th quantile to time $t$.

With this approach, the network is trained to determine the correct quantile given some specific ensemble inputs, $\mathbf{X}_t$, and the true observation, $y_t$. By simultaneously learning for all quantiles, the LSTM network effectively eliminates any crossings in the corrected ensembles, that is, in comparison to fitting a separate network for every quantile. 

\begin{figure}[ht]
    \centering
    \includegraphics[width=1\linewidth]{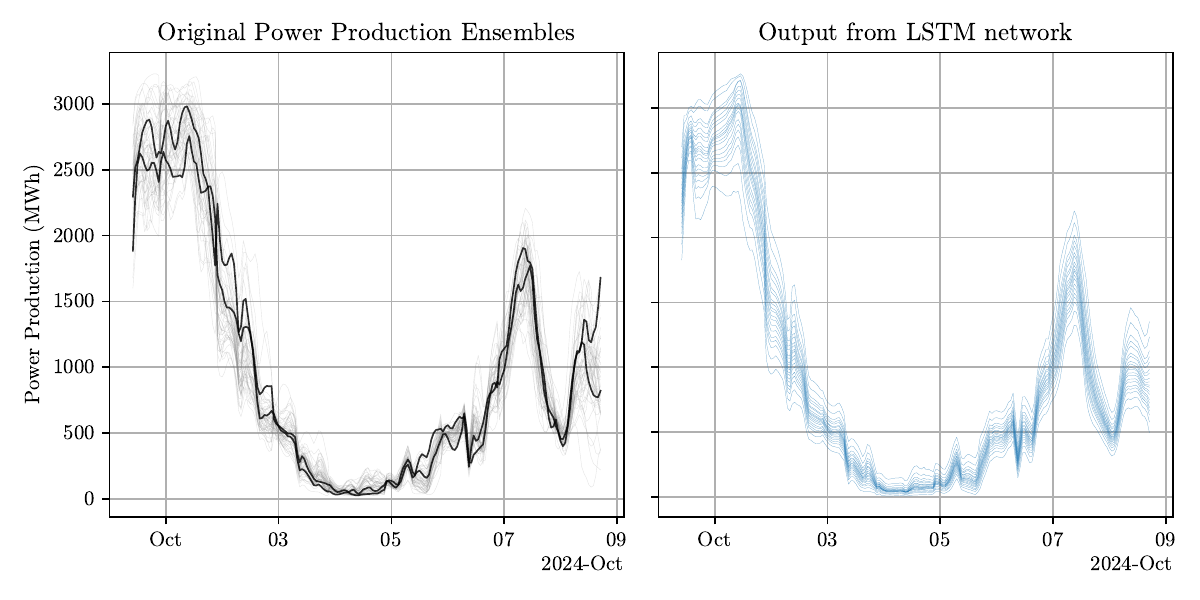}
    \caption{\textit{Left:} Original Power Production Ensembles with two highlighted ensemble members crossing multiple times. \textit{Right:} Output from the LSTM network showing no crossings.}
    \label{fig:outputFromLSTM}
\end{figure}

Fig. \ref{fig:outputFromLSTM} illustrates the output generated by the LSTM network using the previously described loss function. The output from the LSTM network demonstrates an absence of crossings, contributing to the lack of crossings and enhanced reliability in the final output from NABQR.

\subsubsection{Model Selection}
\label{sec:modelsel}

For the LSTM models, we utilized TensorFlow's batch size parameter, which dynamically handles batch sizes. In the LSTM case, the input shape corresponds to a quadruplet with a dimension of units, batch size, number of quantiles, and number of time steps. To maintain conciseness, we omit the training and validation loss curves for all models. They are satisfactory.

Table \ref{tab:lstm_model_selection} shows the model selection process for the LSTM network, where activation functions for Dense 1 and 2 are sigmoid and Rectified Linear Unit (ReLU), respectively. \verb|n_quantiles| is the number of quantiles we input to the LSTM module and \verb|n_out| refers to the number of nodes in the dense layers and the output nodes.

We began the LSTM experiments with Model 1 using lagged values from hour $t-1$ as input. Model 2 extended the input to $(t-1,t-2,t-3)$, but this yielded no significant improvement. To capture more correlations, Model 3 added two dense layers and lagged time steps from the last 12 hours. Model 4 incorporated lagged data from the previous 48 hours and additional neurons, offering a more complex structure but at a high computational cost. To balance performance and efficiency, Model 5 reduced lagged time steps and neurons, achieving comparable results with an acceptable training time.

See Table \ref{tab:lstmnn} for a parameter specification for Model 5. The total number of trainable parameters for Model 5 is 321,976. The primary contribution in parameter count stem from the LSTM module with 256 units.

\begin{table*}[ht]
    \centering
    \begin{tabular}{cccccc}
        \toprule
        \multicolumn{5}{c}{LSTM} \\
        \midrule
        Model & LSTM Layer & \multicolumn{2}{c}{Dense Layers} & Output Layer \\
        & & Dense 1 & Dense 2 & \\
        \midrule
        1 & (\verb|n_quantiles|, 1) &  &  & \verb|n_out| \\
        2 & (\verb|n_quantiles|, [1,2,3]) &  &  & \verb|n_out| \\
        3 & (\verb|n_quantiles|, [1,2,\ldots,12]) & \verb|n_out| & 100 & \verb|n_out| \\
        4 & (\verb|n_quantiles|, [1,2,3,...,48]) & 200 & 200 & \verb|n_out| \\
        5 & (\verb|n_quantiles|, [1,2,3,6,12,24,48]) & \verb|n_out| & \verb|n_out| & \verb|n_out| \\
        \bottomrule
    \end{tabular}
    \caption{LSTM model selection process where \texttt{n\_quantiles} = 51 and \texttt{n\_out} = 20.}
    \label{tab:lstm_model_selection}
\end{table*}

\begin{table}[ht]
\centering
\begin{tabular}{lll}
\toprule
Layer Name & Weight Type & Shape \\
\midrule
lstm & Layer Weights & (51, 1024) \\
lstm & Recurrent Weights  & (256, 1024) \\
lstm & Bias Weights  & (1024,) \\
dense$_1$ & Layer Weights  & (256, 20) \\
dense$_1$ & Bias Weights  & (20,) \\
dense$_2$ & Layer Weights  & (20, 20) \\
dense$_2$ & Bias Weights  & (20,) \\
\bottomrule
\end{tabular}
\caption{Model 5 parameter specification for the LSTM network.}
\label{tab:lstmnn}
\end{table}

\subsection{Time-Adaptive Quantile Regression}
\label{sec:taqr}
We will now explain the second main part of this methodology. 
\subsubsection{Quantile Regression}
Let $Y$ be a real-valued random variable with $F_Y(y)=P(Y \leq y)$. The $\tau$'th quantile of $Y$ is then given by
\begin{align}
q_Y(\tau)=F_Y^{-1}(\tau)=\inf \left\{y: F_Y(y) \geq \tau\right\},
\label{eq:q_def}
\end{align}
where $\tau \in(0,1)$.

Quantile regression differs from Least Squares regression by its loss function. Instead of the usual MSE loss function, for quantile regression, we use the following check mark loss function 
\begin{equation}
L_\tau(m)=m\left(\tau-\mathbb{I}_{(m<0)}\right), 
\label{eq:LossFunc}    
\end{equation}
where $\mathbb{I}$ is an indicator function and $m \in \mathbb{R}$ \citep{Koenker1978, Koenker2005}. If we minimize the expected loss of $Y-f(x)$ with respect to $f(x)$, a specific quantile $\tau$ is found
\begin{align}
q_Y(\tau)&=\underset{f(x)}{\arg \min } \mathbb{E} \left(L_\tau(Y-f(x))\right) \\
\label{eq:qrprob}
& =\underset{f(x)}{\arg \min }\bigg\{(\tau-1) \int_{-\infty}^{f(x)}(y-f(x)) d F_Y(y)\\
&+\tau \int_{f(x)}^{\infty}(y-f(x)) d F_Y(y)\bigg\} .
\end{align}
For properties on quantile regression, we refer to \citep{Koenker2005}. Standard quantile regression can be slow for larger problems with many quantiles and long time periods – especially when an online implementation has to rerun the regression every time new data becomes available. Hence, we now introduce the adaptive aspect of the algorithm.  

\subsubsection{Adding Time Adaptivity}
We introduce key concepts of the TAQR algorithm originally developed in \citep{adaptiveqr1, adaptiveqr2, adaptiveqr3}, and \citep{taqrpaper}. We follow the methodology in \citep{taqrpaper}, which we also refer to for further details.

Let $y$ be the observed data and $\hat{Q}$ the estimated quantile, then the positive and negative part of the residuals, $(y-\hat{Q})$, is given by $\mathbf{r}^{+}= \{ \max(r_i,0) \}_{i=1,...,n}$ and $\mathbf{r}^{-} = \{ \max(-r_i,0) \}_{i=1,...,n}$, respectively. We can then write the LP formulation of the quantile regression problem for the $\tau$'th quantile as \citep{Koenker2005}
\begin{align}
\min_{ \beta \in \mathbb{R}^K} \{\tau \mathbf{1}^T \mathbf{r}^{+}+(1-\tau) &\mathbf{1}^T \mathbf{r}^{-}: \mathbf{X} \beta+\mathbf{r}^{+}-\mathbf{r}^{-}=\mathbf{y},\notag \\
&\left(\mathbf{r}^{+}, \mathbf{r}^{-}\right) \in \mathbb{R}_+^{2 n}, \beta \in \mathbb{R}^K\}.
\label{eq:lpprob}
\end{align}

To solve the minimization problem we employ Dantzig's simplex method \citep{dantzig1954simplex}. We refer to \citep{taqrpaper} for a full explanation of an actual implementation. To add time adaptivity to quantile regression we will solve the LP formulation of the problem. This is a big advantage since we can provide a start guess for the algorithm.

Initially, we solve a small (192 data points) quantile regression problem with the rq-package in R. Here, we obtain $K$ residuals close enough to zero and $K$ parameter estimates, such that we have a feasible start guess to warm start the simplex algorithm. 

For time $t$, we use knowledge about the solution to time $t-1$. This allows us to converge extremely fast and often within two iterations on the polytope spanned by the design matrix. The reason is that the density of data looks very similar when we only replace the oldest data point with one new one. 

We include $N_{\text{TAQR}}=5000$ hours worth of data in the design matrix, $\mathbf{X}$, using a sliding window. This ensures that the parameters in the TAQR algorithm have seen both periods with high and low wind production. This is necessary to get good results. Including fewer data points, e.g., 192 (nine days), results in jagged and extreme predictions from the TAQR algorithm. The reason is that if the algorithm in the last 192 hours only experienced low power production due to calm winds and suddenly the wind power production ramps up, then the algorithm produced results as extreme as 10 or even 100 times the actual observations.

\subsection{NABQR in an Online Setting}
\label{sec:online}
Assuming that the network is seldom retrained, we effectively already have an online implementation of the method, since TAQR works in an online setting. See Section \ref{sec:app_online} for a further discussion.

Let $\mathbf{X_R}$ denote the original ensembles from the forecast provider. Let $f_{LSTM}: (\mathbf{X_R},t) \mapsto \hat{X}_t$ be the LSTM transformation, where $\mathbf{X_R}_{,t} \in \mathbb{R}^{51 \times |AR|}$, and $|AR|$ is the number of timesteps we include in the lookback (auto-regressive) part of the LSTM network.

Furthermore, $\hat{{X}}_t \in \mathbb{R}^{20}$, where the dimensionality of 20 was selected experimentally to prevent certain corrected ensembles from becoming identically zero.

In the implementation presented here, $AR=[1,2,3,6,12,24,48 ]$ and thus $|AR|=7$. Let $\hat{\mathbf{X}}_{t-N_{\text{TAQR}},...,t} \in \mathbb{R}^{20 \times N_{\text{TAQR}}}$ be the design matrix for the TAQR algorithm and $g_{TAQR}: (\hat{\mathbf{X}}_{t-N_{\text{TAQR}},...,t},\tau) \mapsto \hat{q}_{\tau,t} \in \mathbb{R}^{24}$ be the linear mapping the TAQR algorithm does. $\hat{q}_{\tau,t}$ is in $\mathbb{R}^{24}$ since for every 24 hours we obtain 24 hourly day-ahead forecasts. 

Thus, the method makes the following prediction
\begin{equation}
    \hat{q}(\tau, t) = g_{TAQR}(f_{LSTM} (\mathbf{X_R},t),\tau). 
\end{equation}

\begin{figure}
    \centering
    \includegraphics[width=1\linewidth]{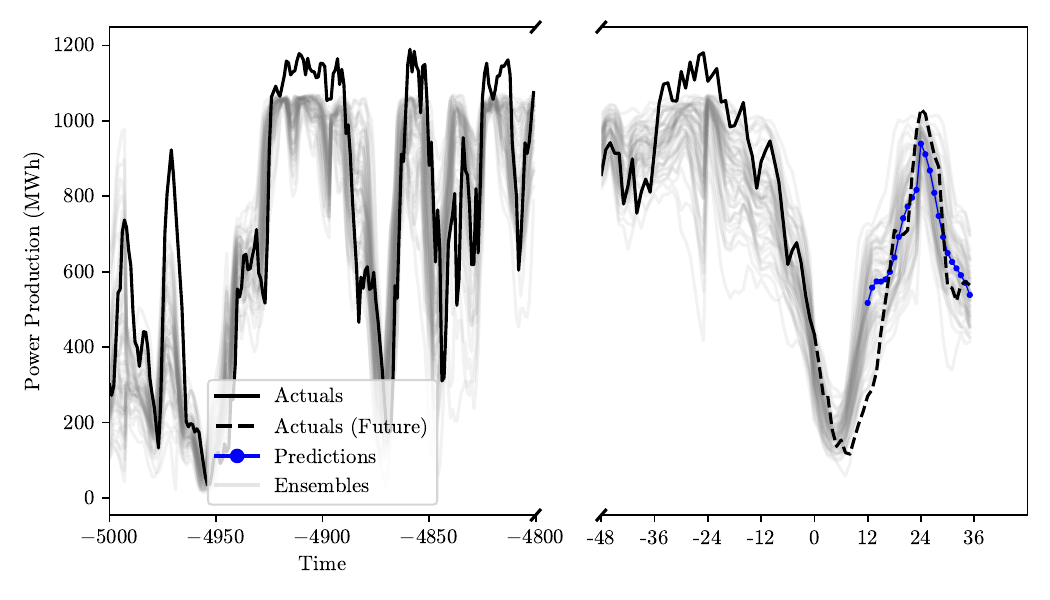}
    \caption{
    Example of data, including ensembles and median prediction 12-36 hours ahead.
    }
    \label{fig:timeoverview}
\end{figure}

Once new ensembles become available we update the design matrix elements with the newest corrected ensembles. Fig.~\ref{fig:timeoverview} shows an example of a snapshot in time. The TAQR algorithm uses data up to 5000 data points back in time, while the LSTM network uses the AR time-indexes. The ensembles are available into the future $0<t<36$, while the actual observations are not, and thus shown with a dashed line. We illustrate only our median day-ahead predictions for clarity.

\subsection{Quantile Random Forest and Quantile Gradient Boosting}

\paragraph{Quantile Random Forest (QRF)}
QRF \citep{qrfMein2006} is an adaptation of random forests designed to estimate the entire conditional distribution of \(y\mid \mathbf{X}\), from which any quantile can be extracted. Recall that in a standard random forest, each terminal node stores and averages the \(\{y_i\}\) values falling into that node, producing an estimate of the conditional mean. In contrast, QRF retains all observed \(\{y_i\}\) values in each terminal node.

Let \(\mathcal{D} = \{(\mathbf{X}_i, y_i)\}_{i=1}^{N_{\text{train}}}\) be the training dataset, as in Table \ref{tab:traintestset} where \(\mathbf{X} \in \mathbb{R}^{51\times N_{\text{train}}} \) and \(y \in \mathbb{R}^{N_{\text{train}}} \). QRF grows an ensemble of decision trees, each constructed from a bootstrap sample of \(\mathcal{D}\). For a test point \(x\), each tree assigns weights \(w_i(x,\theta)\) to sample points \((\mathbf{X}_i, y_i)\), usually by checking if \(\mathbf{X}_i\) ends in the same terminal node as \(x\). The overall forest weight is the average,
\[
  w_i(x) \;=\; \frac{1}{M} \sum_{\theta=1}^{M} w_i(x,\theta),
\]
where \(M\) is the number of trees. Let
\[
  F_n(y \mid x) \;=\; \sum_{i=1}^{N} w_i(x)\,\mathbf{1}\bigl(y_i \leq y\bigr).
\]
Then \(F_n(\cdot \mid x)\) is an empirical distribution function over the weighted set of training responses. Any conditional quantile is obtained by
\[
   \hat{Q}_\tau(x) \;=\; \inf\Bigl\{y \,\big\vert\, F_n\bigl(y \mid x\bigr) \ge \tau\Bigr\}.
\]
Hence, QRF provides a nonparametric estimate of the entire conditional distribution of \(y\), from which one directly extracts arbitrary quantiles \(\hat{Q}_\tau(x)\).

\paragraph{Quantile Gradient Boosting (QGB)}
Quantile Gradient Boosting \citep{friedman1999stochastic, friedman2001greedy} generalizes the idea of additive modeling and boosting to optimize the check loss \eqref{eq:LossFunc} for a desired \(\tau\). In ordinary gradient boosting, one fits an additive function
\begin{equation}
    F_m(\mathbf{X}) \;=\; F_{m-1}(\mathbf{X}) \;+\; \rho_m\,h_m(\mathbf{X}),
\end{equation}
where each weak learner \(h_m\) is trained to fit the negative gradients (pseudo-residuals) of the current loss. For quantile regression, the pseudo-residual for sample \(i\) at iteration \(m\) becomes
\begin{align*}
    r_i^{(m)} & =\; \frac{\partial L_{\tau}(y_i, F_{m-1}(\mathbf{X}_i))}{\partial F_{m-1}(\mathbf{X}_i)} \\
    & =
    \begin{cases}
        \tau, & y_i > F_{m-1}(\mathbf{X}_i),\\
        \tau - 1, & y_i \le F_{m-1}(\mathbf{X}_i).
    \end{cases}
\end{align*}
The weak learner \(h_m\) is then fit by regressing these pseudo-residuals on the features \(\{\mathbf{X}_i\}\). A step size \(\rho_m\) is chosen via line search,

\begin{equation}
    \rho_m \;=\; \arg\min_{\rho} \,\sum_{i=1}^{N} L_{\tau}\bigl(y_i,\; F_{m-1}(\mathbf{X}_i) \,+\, \rho\,h_m(\mathbf{X}_i)\bigr).
\end{equation}

Repeatedly updating the model in this way yields a sequence \(\{F_m\}\) that converges to an estimator of the desired conditional quantile function. In practice, each \(h_m\) can be a tree whose terminal nodes store a constant update. Consequently, QGB is a robust algorithm that iteratively refines a quantile estimate while naturally handling large or complex feature spaces.

\subsection{Error Measures}
\label{sec:errormeasures}
We use the Continuous Ranked Probability Score (CRPS), and Quantile Score (QS) to evaluate the forecasts in a probabilistic setting. Furthermore, the Mean Absolute Error (MAE) score is shown when relevant since this score is often of interest to energy traders. The CRPS evaluates the forecast distribution but does not capture the correlation structure for multivariate forecasts. We use the reliability plot to check that the mean and variance of our forecast ensembles are properly calibrated, see Fig.~\ref{fig:relplot} in Section \ref{sec:discussion}. 

Together, the MAE, Quantile Score, and CRPS score properly allow us to compare our forecasts and determine which are most accurate.

At time step $t$ we have predictions $\hat{y}_{t+12},...,\hat{y}_{t+36}$. Thus, for every day (24 hours), we append the predictions to the vector $\hat{y}$, such that we in the end have one vector containing all predictions for all time steps. 

\textit{Mean Absolute Error (MAE)}\\
The MAE is then defined as the mean of the absolute error:
$$
MAE = \frac{1}{N} \sum_{t=1}^N |y_t - \hat{y}_t|.
$$

For the original and corrected ensembles, $\hat{y}$ is the median ensemble. For the TAQR predictions  $\hat{y}_t = \hat{q}(0.5,t)$. 

\textit{Continuous Ranked Probability Score (CRPS)} \\
For each time step, we calculate the empirical cumulative distribution function $F_t$ based on the ensemble values of interest (original, QRNN, or NABQR). In this article, we are working with the empirical cumulative distribution function calculated based on the values of the sorted quantiles at each time step.
The CRPS for one observation, $y_t$, is defined as
\begin{align}
\label{eq:crps1}
CRPS(F_t,y_t) = \int^\infty_\infty (F_t(u)- \mathbbm{1}(u \geq y_t))^2 du
\end{align}

We are interested in the average (over all observations $y$). When presenting the score, we will only write "CRPS".

Often it is a good idea to split up the integral into the following two parts
\begin{equation}
\label{eq:crps2}
\operatorname{CRPS}(F_t, y_t)=\int_{-\infty}^{y_t} F_t(u)^2 \mathrm{~d} u+\int_{y_t}^{\infty}(F_t(u)-1)^2 \mathrm{~d} u.
\end{equation}

We will use the \verb|crps_ensemble| function from the \verb|properscoring| module to calculate our CRPS score \citep{properscoring}. This module uses a vector-optimized version of \eqref{eq:crps2}. As a concrete example, consider the CRPS score in Table \ref{tab:performance_metrics_combined}.a for NABQR: here the empirical cumulative distribution function $F_t$ is estimated from the predicted 13 quantiles from the full NABQR methodology.

We will define the \textit{Quantile Score} (QS) as the check mark loss function \eqref{eq:LossFunc} applied to our predictions compared to the actual observations.
The Quantile Score for the quantile $\tau$ is defined as
    $$
    QS_\tau(y,\hat{y}) = \frac{1}{N} \sum_{i=t}^N L_\tau(y_t - \hat{y}_t),
    $$
    where $L_\tau$ is defined in \eqref{eq:LossFunc}. We use the average (over all estimated quantiles, $\boldsymbol{\tau}$) quantile score $\overline{QS}=\frac{1}{|\boldsymbol{\tau}|}\sum_{\tau \in \boldsymbol{\tau}} QS_{\tau}(y,\hat{y})$.

\section{Application to Forecasting of Wind Power Production in Denmark}
\label{sec:results}

\subsection{Reliability} \label{sec:reliability}
To be able to properly evaluate our forecasts, we need calibrated forecasts \citep{mathiasAI}. To evaluate whether the proposed method better captures the nominal quantiles and check if the forecasts are calibrated, we will use a reliability plot. In a reliability plot, we compare the observed frequencies to the nominal quantiles. 

\begin{figure}
    \centering
    \includegraphics[width=1\linewidth]{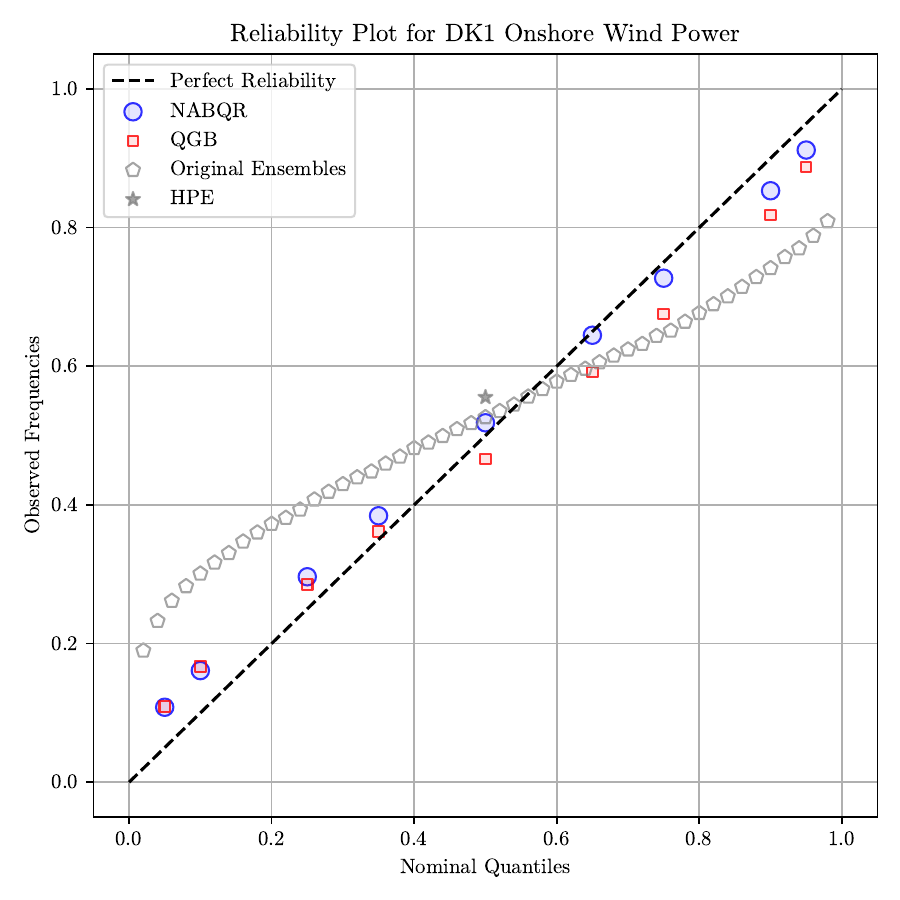}
    \caption{Reliability plot for DK1 Onshore Wind Power showing the original ensembles compared to predictions from TAQR, QGB and NABQR.}
    \label{fig:rel_DK1on}
\end{figure}

Fig. \ref{fig:rel_DK1on} presents the reliability plot for DK1 onshore wind power. For the original ensembles, the lower quantiles tend to be overshot, indicating insufficient ensemble representation in the lower data range, while the higher quantiles exhibit the opposite behavior. This suggests the original ensembles are too clustered around the median, confirming that they require correction. By design, TAQR should improve the reliability, since it models the quantiles directly. Furthermore, we specifically optimize for the quantile loss function once we correct the ensembles. Consequently, as seen in Fig. \ref{fig:rel_DK1on}, employing TAQR and NABQR enhances the reliability, aligning observed frequencies more closely with nominal quantiles. 

HPE refers to the high probability ensembles, the unperturbed control ensemble from Section \ref{sec:data}, which can be seen at the nominal median and an observed frequency of approximately 0.58 (above the blue dot). Thus, the HPE is not properly calibrated and is overshooting the produced power on average. 

The original ensembles are poorly calibrated, which makes sense, given they are not created as quantiles, while the NABQR and QGB forecasts are. They are very close to perfect reliability. If one is only interested in the best reliability, one should use the NABQR predictions. The general tendency for all methods is to have a small overshoot for lower quantiles and a slight undershoot for higher quantiles.

For the other areas, we see similar improvements as for DK1 onshore wind power production (\ref{sec:appendix_rel}, Fig. \ref{fig:relplot}).
 
\subsection{In-sample vs.~out-of-sample effect of correcting ensembles}
\label{sec:isvoos}
\begin{table*}[ht]
    \centering
    
    \begin{tabular}{lrrrr}
    \toprule
    Metric / Wind area & DK1-On & DK1-Off & DK2-On & DK2-Off \\
    \midrule
    CRPS In-Sample & 1.010 & 0.977 & 0.910 & 0.958 \\
    CRPS Out-of-Sample & 1.000 & 0.995 & 0.916 & 0.964 \\
    QS In-Sample & 0.976 & 0.978 & 0.895 & 0.936 \\
    QS Out-of-Sample & 0.966 & 0.974 & 0.896 & 0.943 \\
    MAE In-Sample & 1.004 & 1.048 & 0.961 & 1.001 \\
    MAE Out-of-Sample & 0.988 & 1.000 & 0.956 & 1.000 \\
    \bottomrule
    \end{tabular}
    \caption{"On" refers to Onshore and "Off" refers to Offshore power production. We compare the in- and out-of-sample performance of the LSTM corrected vs.~original ensembles. $<1$ indicates corrected ensembles are better.}
\label{tab:is_v_oos}
\end{table*}

In Table \ref{tab:is_v_oos}, we present a comparison between the in- and out-of-sample performance of the corrected vs.~original ensembles. The relative score, $RS$, which we calculate both in sample and out of sample, is presented as $RS=\frac{S_{Corrected}}{S_{Raw}}$ or in a more general notation: 
\begin{equation}
    RS=\frac{S_{M}}{S_{B}}
\end{equation}
where $RS$ is the relative score, $S_B$ is the baseline score, and $S_M$ is the score for the model we consider. $RS<1$ indicates improvement. 

Since our multidimensional loss function used to train the LSTM is directly related to the QS score presented in Section \ref{sec:errormeasures}, we expect this to be better in sample. In general, the corrected ensembles perform better both in- and out-of-sample compared to the original ensembles. The MAE is close to 1 in any case. We will only see improvements in MAE once TAQR is applied as the second step in NABQR.

\subsection{Results of NABQR predictions} 
\label{sec:taqrpred}
To calculate the quantile score we need to specify which quantiles are used. It is not clear what this range is for the original ensembles, since they are not quantiles. We assume that the 51 ensembles span the range from 5\% to 95\%, equidistantly. Changing this to the range from 1\% to 99\% does not affect the results noticeably. For the LSTM corrected ensembles we have 20 equidistant ensembles from 5\% to 95\%. The marginal TAQR predictions consist of the following 13 quantiles: [0.05, 0.1, 0.15, 0.25, 0.35, 0.45, 0.5, 0.55, 0.65, 0.75, 0.85, 0.9, 0.95].

To demonstrate the performance gain of NABQR, we compare it with standard quantile regression, QRF, and the state-of-the-art method QGB as motivated in Section \ref{intro}. The QGB is better on all measures than standard quantile regression and QRF, thus this is the only score we report in Table \ref{tab:performance_metrics_combined}. We create QGB with 50 estimators, a learning rate of 0.1, and a max depth of 3 in the scikit-learn environment \citep{scikitlearn}. We further note that the QRNN from the first step in NABQR could also be considered as the corrected ensembles for the basis in TAQR.

Table \ref{tab:performance_metrics_combined} shows results for wind power production in DK1 and DK2. In the first row, we present the actual absolute score for the original ensembles, which we will consider the baseline. The others are calculated as the relative score, $RS$, introduced in Section \ref{sec:isvoos}.

\begin{table*}[ht]
\centering
\begin{subtable}[t]{0.48\textwidth}
    \caption{DK1 Offshore Wind Power}
    \centering
    \resizebox{\textwidth}{!}{\begin{tabular}{lccc}
    \toprule
      & MAE & CRPS & QS \\
    \bottomrule
    Original Ensembles & 103.170 & 78.917 & 41.616 \\
    \toprule
    QRNN & 0.976 & 0.974 & 0.949 \\
    TAQR & \textbf{0.939} & \textbf{0.904} & \textbf{0.841} \\
    QGB & 1.110 & 0.953 & 0.905 \\
    NABQR & 0.961 & 0.915 & 0.845 \\
    \bottomrule
    \end{tabular} }
\end{subtable}%
\hfill
\begin{subtable}[t]{0.48\textwidth}
    \caption{DK1 Onshore Wind Power}
    \centering
    \resizebox{\textwidth}{!}{\begin{tabular}{lccc}
    \toprule
      & MAE & CRPS & QS \\
    \bottomrule
    Original Ensembles & 208.693 & 156.171 & 82.895 \\
    \toprule
    QRNN & 0.995 & 1.036 & 0.997 \\
    TAQR & 1.028 & 1.004 & 0.959 \\
    QGB &           0.987 & 0.953 & 0.816 \\
    NABQR & \textbf{0.911} & \textbf{0.877} & \textbf{0.806} \\
    \bottomrule
    \end{tabular} }
\end{subtable}

\vspace{0.3cm} 

\begin{subtable}[t]{0.48\textwidth}
    \caption{DK2 Offshore Wind Power}
    \centering
    \resizebox{\textwidth}{!}{\begin{tabular}{lccc}
    \toprule
     & MAE & CRPS & QS \\
    \bottomrule
    Original Ensembles & 117.970 & 88.176 & 46.928 \\
    \toprule
    QRNN & 0.948 & 0.979 & 0.950 \\
    TAQR & 0.946 & 0.908 & 0.844 \\
    QGB & 0.933 & 0.911 & \textbf{0.788} \\
    NABQR & \textbf{0.930} & \textbf{0.887} & {0.811} \\
    \bottomrule
    \end{tabular} }
\end{subtable}%
\hfill
\begin{subtable}[t]{0.48\textwidth}
    \caption{DK2 Onshore Wind Power}
    \centering
    \resizebox{\textwidth}{!}{\begin{tabular}{lccc}
    \toprule
      & MAE & CRPS & QS \\
    \bottomrule
    Original Ensembles & 62.761 & 48.111 & 25.308 \\
    \toprule
    QRNN & 0.976 & 0.972 & 0.943 \\
    TAQR & 0.605 & 0.576 & 0.534 \\
    QGB & 0.602 & 0.574 & \textbf{0.496} \\
    NABQR & \textbf{0.602} & \textbf{0.565} & {0.523} \\
    \bottomrule
    \end{tabular} }
\end{subtable}
\caption{Performance metrics for different methods on DK1 and DK2 wind power production. QRNN refers to the corrected ensembles after the first step in NABQR. All results are out-of-sample.}
\label{tab:performance_metrics_combined}
\end{table*}

The method generally seems to be performing well. 
We see good improvements across all four scores for DK1 onshore wind power production and even bigger improvements for DK2 onshore wind power production. 

Applying TAQR to the corrected ensembles (QRNN) seems to be a good idea, in three of the four areas. The NABQR results are close to the TAQR results for DK1 offshore, indicating that the choice of method should be investigated further.

In general, we believe that the superior performance obtained for onshore compared to offshore wind is due to the curtailment of offshore wind power plants as we will discuss in Section \ref{sec:discussion}.

\begin{figure}
    \centering
    \includegraphics[width=1\linewidth]{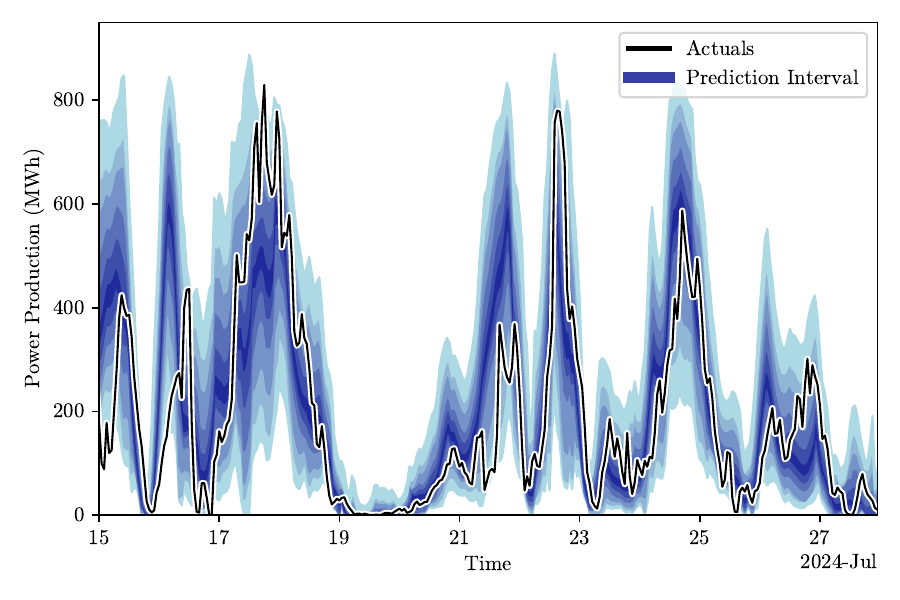}
    \caption{Excerpt of NABQR out-of-sample predictions with quantile interval 5\%-95\% shown in shades of blue for DK2 offshore wind power production.}
    \label{fig:example_dk2_off}
\end{figure}

Figure \ref{fig:example_dk2_off} illustrates the TAQR prediction intervals and the actual observations for a short period in July 2024.

\section{Application to Energy Trading} \label{sec:applications}
We present a straightforward application of our methodology, which is based on strong assumptions. These assumptions are introduced to maintain conciseness in this section.

In this case study, we adopt the perspective of an energy trader. If our median from the NABQR forecasts suggests a higher value than the original ensemble median, this signals an anticipated increase in supply. Based on fundamental supply-demand principles, such an increase would likely drive imbalance prices for the following day lower than the spot prices. Imbalance price is the price the TSO pays producers/consumers to up- or down regulate their production/consumption to ensure grid stability. Note, the imbalance price can be negative.

Assuming these supply-demand dynamics hold, we implement a trading strategy by selling electricity in the day-ahead spot market and repurchasing it at a lower imbalance price the next day. This approach effectively shorts the market, and we aim to evaluate its profitability under varying conditions.

Conversely, if our forecasts indicate lower production, implying a supply decrease, we would purchase electricity in the day-ahead market and sell it at a potentially higher imbalance price the next day. This case study explores whether such a strategy is profitable.

The pseudocode for this trading setup can be seen in Algorithm \ref{algo:trading_strategy}. Note that the comparison (line 5) is adjusted by an offset, $o$, which is the mean error of the difference between the predictions and actual observations in the training set.

\begin{algorithm}
\caption{Trading Strategy}
\begin{algorithmic}[1]
\State Predictions: $\hat{y}$, $\hat{y}_{0.5}$ is the median 
\State Original Ensembles: $y$ 
\State Mean Offset Based on Training Set: $o$
\State Index set for the test set: $I$
\For{$i \in I$}
    \If{$\hat{y}_{0.5}[i]-o_i > {y}_{0.5}[i]$}
        \State \textit{Sell Spot, Buy Imbalance}
    \Else
        \State \textit{Buy Spot, Sell Imbalance}
    \EndIf
    \State \textit{Update Profit and Loss}
\EndFor
\State \textbf{Return} \textit{Total Profit and Loss}
\end{algorithmic}
\label{algo:trading_strategy}
\end{algorithm}

\begin{figure}
    \centering
    \includegraphics[width=1\linewidth]{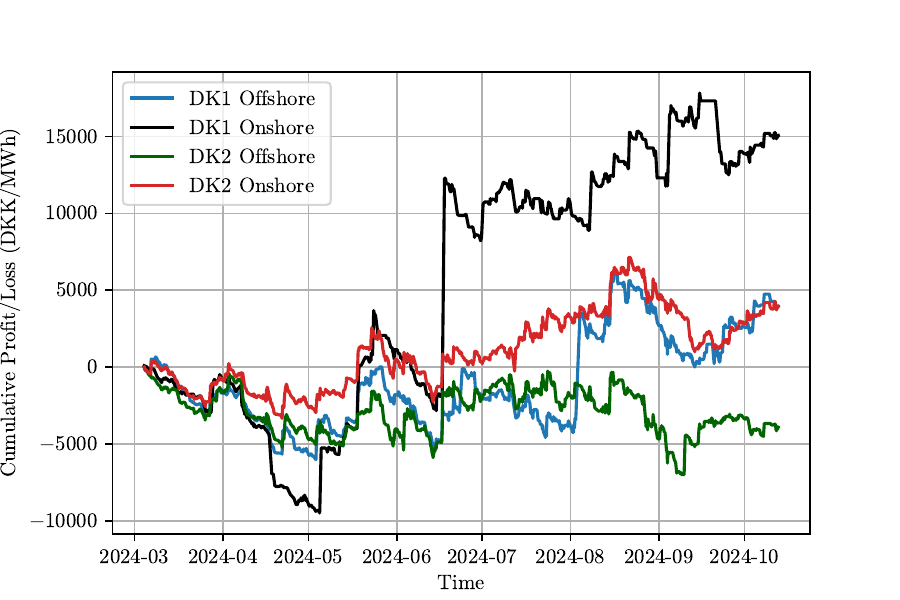}
    \caption{Cumulative profit and loss in DKK for trading size 1 MWh per trade.}
    \label{fig:pnlfig}
\end{figure}

In Fig.~\ref{fig:pnlfig}, we show the cumulative profit and loss for the trading strategy described in Algorithm \ref{algo:trading_strategy} when trading 1 MWh per trade. We see that DK1 Onshore wind power is the greatest performer, but it is also the area where the most power is produced (see Fig.~\ref{fig:densityplot}). It makes sense that DK1 Onshore is the best performer, since this is the biggest driver of price changes, given its volume relative to the entire DK1 power production. Across the four areas, the four strategies collectively generate 19,041 DKK with a trading size of 1 MWh per trade over the test set.

\section{Discussion and Future Work}
\label{sec:discussion}
\subsection{Correlation Structure} 
The process of generating power production ensembles involves perturbing initial conditions and propagating them forward using deterministic PDE systems. This gives the ensembles a positive correlation with the actual observations and significant positive autocorrelation for more than 100 lags, see \citep{bastianmsc} for a further discussion of the autocorrelation. However, when these ensembles are modified, the intrinsic autocorrelation structure within the data is often disrupted. To address this issue, future work should consider restoring the autocorrelation structure in the residuals by employing stochastic differential equations, e.g., following the approach outlined in \citep{sde1}.

\subsection{Online Implementation}
\label{sec:app_online}
In a production setting, new observations are available every 5 minutes. The forecasted power production ensembles are updated every 6 hours. This allows for intra-day modeling of a quantile estimate of the forecast distribution, giving even more merit to the use of TAQR. 
Increasing the frequency of forecasts brings many challenges. One is retraining the neural networks: TensorFlow and PyTorch provide the possibility of warm-starting your training from a previous training period. Switch-EMA is another method for retraining neural networks \citep{li2024switchemafreelunch}. Since TAQR is already adaptive, we do not need to retrain the neural networks more often than, e.g.~every seventh day. 

As renewable energy production continues to expand, it is important to retrain the network once new production capacity is installed and sufficient data including the new capacity becomes available. This approach provides a more effective retraining schedule than relying on fixed calendar-based intervals.

\subsection{Further Areas of Application: Solar}
In Denmark, solar power is growing even faster than wind power. Solar power comes with its own set of challenges. For example, how to deal with the nighttime, where no power is produced. Power production ensembles are available for solar power and are an area of interest, where NABQR could potentially improve forecasts as well.

\subsection{Modeling of Cost Function}
To enhance the trading strategy, one should account for the covariation of returns, the correlation structure among trading strategies, and their respective time horizons. A natural step would be to model the cost function of bidding in energy markets given the improved quantile forecasts in this article \citep{finnah2022optimal}.

\section{Conclusion} \label{sec:conclusion}
In this article, we have presented a novel approach to using neural networks as a data-driven basis for TAQR. LSTM networks were found to be the best neural network architecture. We used the LSTM network to correct power production ensembles based on a multidimensional quantile loss function.  We included 5000 data points in the design matrix for the TAQR algorithm in a sliding window. 

We applied the methodology to wind power production in Denmark. The method provided improvements quantified by MAE, CRPS, and QS. The method performed best on onshore wind power with improvements of up to 48\% in the QS score. In most cases, NABQR beats state-of-the-art methods such as Quantile Gradient Boosting.

We considered a simple trading strategy as an application of the improved forecasts for energy traders in the day-ahead and imbalance market. The strategy was profitable in three of the four areas; however, the returns were only significantly greater than zero in one case. 

Looking ahead, NABQR's framework can be extended to other renewable energy sources, such as solar power, which presents its own unique forecasting challenges. Additionally, future work could explore the integration of stochastic differential equations to restore the autocorrelation structure in residuals and treat the challenges that arise with an online implementation.

In conclusion, NABQR demonstrates significant potential in renewable energy forecasting, providing a robust tool for enhancing the accuracy and reliability of probabilistic wind power forecasts. Its user-friendly, open-source Python package is freely accessible, making it a valuable resource for the field.

\subsection*{CRediT authorship contribution statement}
\textbf{Bastian Schmidt Jørgensen:} Writing– original draft, Project administration, Investigation, Data curation, Software, Methodology. 

\textbf{Jan Kloppenborg Møller:} Writing– review \& editing, Conceptualization, Software. 

\textbf{Peter Nystrup}:
Writing– review \& editing, Data curation, Conceptualization. 

\textbf{Henrik Madsen:} Writing– review \& editing, Conceptualization, Funding, Project administration.

\subsection*{Declaration of competing interest}
The authors declare that they have no known competing financial
interests or personal relationships that could have appeared to influence the work reported in this article.
\subsection*{Data availability}
The authors do not have permission to share data, but simulated data is available in the Python package \texttt{nabqr}.
\section*{Acknowledgments}

This work is supported by \textit{IIRES} (Energy Cluster Denmark),  \textit{SEM4Cities} (Innovation Fund Denmark, No. 0143-0004), \textit{ELEXIA} (Horizon Europe No. 101075656), and the projects \textit{DynFlex} and \textit{PtXMarkets}, which both are part of the Danish Mission Green Fuel portfolio of projects. 

\newpage
\bibliographystyle{elsarticle-num-names} 
\bibliography{sample}

\begin{thebibliography}{50}
\expandafter\ifx\csname natexlab\endcsname\relax\def\natexlab#1{#1}\fi
\providecommand{\url}[1]{\texttt{#1}}
\providecommand{\href}[2]{#2}
\providecommand{\path}[1]{#1}
\providecommand{\DOIprefix}{doi:}
\providecommand{\ArXivprefix}{arXiv:}
\providecommand{\URLprefix}{URL: }
\providecommand{\Pubmedprefix}{pmid:}
\providecommand{\doi}[1]{\href{http://dx.doi.org/#1}{\path{#1}}}
\providecommand{\Pubmed}[1]{\href{pmid:#1}{\path{#1}}}
\providecommand{\bibinfo}[2]{#2}
\ifx\xfnm\relax \def\xfnm[#1]{\unskip,\space#1}\fi
\bibitem[{{Energistyrelsen}(2024)}]{energistyrelsen2024}
\bibinfo{author}{{Energistyrelsen}}, \bibinfo{title}{Energistatistik 2023}, \bibinfo{publisher}{Energistyrelsen}, \bibinfo{address}{Copenhagen, Denmark}, \bibinfo{year}{2024}. \URLprefix \url{http://www.ens.dk}, \bibinfo{note}{data, tabeller, statistikker og kort om energiforbrug, emissioner og beregningsforudsætninger for perioden 1972-2023.}
\bibitem[{Jha et~al.(2020)Jha, Lee, Iyengar, Hajiesmaili, Irwin, and Shenoy}]{lessfossil0}
\bibinfo{author}{R.~Jha}, \bibinfo{author}{S.~Lee}, \bibinfo{author}{S.~Iyengar}, \bibinfo{author}{M.~H. Hajiesmaili}, \bibinfo{author}{D.~Irwin}, \bibinfo{author}{P.~Shenoy},
\newblock \bibinfo{title}{Emission-aware energy storage scheduling for a greener grid},
\newblock in: \bibinfo{booktitle}{Proceedings of the Eleventh ACM International Conference on Future Energy Systems}, e-Energy ’20, \bibinfo{publisher}{ACM}, \bibinfo{year}{2020}, p. \bibinfo{pages}{363–373}. \DOIprefix\doi{10.1145/3396851.3397755}.
\bibitem[{Madsen et~al.(2025)Madsen, Tohidi, Ebrahimy, Banaei, Ritschel, and Mahdavi}]{madsen2025a}
\bibinfo{author}{H.~Madsen}, \bibinfo{author}{S.~S. Tohidi}, \bibinfo{author}{R.~Ebrahimy}, \bibinfo{author}{M.~Banaei}, \bibinfo{author}{T.~K. Ritschel}, \bibinfo{author}{N.~Mahdavi},
\newblock \bibinfo{title}{Recent trends in demand-side flexibility},
\newblock \bibinfo{journal}{Lecture Notes in Computer Science}  (\bibinfo{year}{2025}) \bibinfo{pages}{167--184}. \DOIprefix\doi{10.1007/978-3-031-74741-0_12}.
\bibitem[{Dahlgren et~al.(2003)Dahlgren, Liu, and Lawarr{\'e}e}]{dahlgren2003risk}
\bibinfo{author}{R.~Dahlgren}, \bibinfo{author}{C.-C. Liu}, \bibinfo{author}{J.~Lawarr{\'e}e},
\newblock \bibinfo{title}{Risk assessment in energy trading},
\newblock \bibinfo{journal}{IEEE Transactions on Power Systems} \bibinfo{volume}{18} (\bibinfo{year}{2003}) \bibinfo{pages}{503--511}.
\bibitem[{Maciejowska et~al.(2024)Maciejowska, Serafin, and Uniejewski}]{costfunc_maciejowska2024}
\bibinfo{author}{K.~Maciejowska}, \bibinfo{author}{T.~Serafin}, \bibinfo{author}{B.~Uniejewski},
\newblock \bibinfo{title}{Probabilistic forecasting with a hybrid factor-qra approach: Application to electricity trading},
\newblock \bibinfo{journal}{Electric Power Systems Research} \bibinfo{volume}{234} (\bibinfo{year}{2024}) \bibinfo{pages}{110541}.
\bibitem[{Wang et~al.(2024)Wang, Zhou, Zhang, Lin, and Wang}]{cVar2023}
\bibinfo{author}{J.~Wang}, \bibinfo{author}{Y.~Zhou}, \bibinfo{author}{Y.~Zhang}, \bibinfo{author}{F.~Lin}, \bibinfo{author}{J.~Wang},
\newblock \bibinfo{title}{Risk-averse optimal combining forecasts for renewable energy trading under cvar assessment of forecast errors},
\newblock \bibinfo{journal}{IEEE Transactions on Power Systems} \bibinfo{volume}{39} (\bibinfo{year}{2024}) \bibinfo{pages}{2296--2309}. \DOIprefix\doi{10.1109/TPWRS.2023.3268337}.
\bibitem[{{ECMWF}(2024)}]{ecmwf_website}
\bibinfo{author}{{ECMWF}}, \bibinfo{title}{Ecmwf: European centre for medium-range weather forecasts}, \bibinfo{howpublished}{\url{https://www.ecmwf.int}}, \bibinfo{year}{2024}. \bibinfo{note}{Accessed: 2024-07-01}.
\bibitem[{Xu et~al.(2023)Xu, Sun, Du, and ce~Gao}]{qr_SOTA}
\bibinfo{author}{C.~Xu}, \bibinfo{author}{Y.~Sun}, \bibinfo{author}{A.~Du}, \bibinfo{author}{D.~ce~Gao},
\newblock \bibinfo{title}{Quantile regression based probabilistic forecasting of renewable energy generation and building electrical load: A state of the art review},
\newblock \bibinfo{journal}{Journal of Building Engineering} \bibinfo{volume}{79} (\bibinfo{year}{2023}) \bibinfo{pages}{107772}. \DOIprefix\doi{https://doi.org/10.1016/j.jobe.2023.107772}.
\bibitem[{Verbois et~al.(2018)Verbois, Rusydi, and Thiery}]{qgbsolar}
\bibinfo{author}{H.~Verbois}, \bibinfo{author}{A.~Rusydi}, \bibinfo{author}{A.~Thiery},
\newblock \bibinfo{title}{Probabilistic forecasting of day-ahead solar irradiance using quantile gradient boosting},
\newblock \bibinfo{journal}{Solar Energy} \bibinfo{volume}{173} (\bibinfo{year}{2018}) \bibinfo{pages}{313--327}. \DOIprefix\doi{https://doi.org/10.1016/j.solener.2018.07.071}.
\bibitem[{Møller et~al.(2008)Møller, Nielsen, and Madsen}]{taqrpaper}
\bibinfo{author}{J.~K. Møller}, \bibinfo{author}{H.~A. Nielsen}, \bibinfo{author}{H.~Madsen},
\newblock \bibinfo{title}{Time-adaptive quantile regression},
\newblock \bibinfo{journal}{Computational Statistics \& Data Analysis} \bibinfo{volume}{52} (\bibinfo{year}{2008}) \bibinfo{pages}{1292--1303}. \DOIprefix\doi{https://doi.org/10.1016/j.csda.2007.06.027}.
\bibitem[{Nielsen et~al.(2006{\natexlab{a}})Nielsen, Nielsen, Madsen, Giebel, Badger, Landberg, Sattler, Voulund, and Tofting}]{windens}
\bibinfo{author}{H.~A. Nielsen}, \bibinfo{author}{T.~S. Nielsen}, \bibinfo{author}{H.~Madsen}, \bibinfo{author}{G.~Giebel}, \bibinfo{author}{J.~Badger}, \bibinfo{author}{L.~Landberg}, \bibinfo{author}{K.~Sattler}, \bibinfo{author}{L.~Voulund}, \bibinfo{author}{J.~Tofting},
\newblock \bibinfo{title}{From wind ensembles to probabilistic information about future wind power production -- results from an actual application},
\newblock in: \bibinfo{booktitle}{2006 International Conference on Probabilistic Methods Applied to Power Systems}, \bibinfo{year}{2006}{\natexlab{a}}, pp. \bibinfo{pages}{1--8}. \DOIprefix\doi{10.1109/PMAPS.2006.360289}.
\bibitem[{Nielsen et~al.(2006{\natexlab{b}})Nielsen, Madsen, and Torben}]{qrExtend}
\bibinfo{author}{H.~A. Nielsen}, \bibinfo{author}{H.~Madsen}, \bibinfo{author}{S.~Torben},
\newblock \bibinfo{title}{Using quantile regression to extend an existing wind power forecasting system with probabilistic forecasts},
\newblock \bibinfo{journal}{Wind Energy} \bibinfo{volume}{9} (\bibinfo{year}{2006}{\natexlab{b}}) \bibinfo{pages}{95--108}. \DOIprefix\doi{https://doi.org/10.1002/we.180}.
\bibitem[{Lauret et~al.(2017)Lauret, David, and Pedro}]{lauretProbsolar}
\bibinfo{author}{P.~Lauret}, \bibinfo{author}{M.~David}, \bibinfo{author}{H.~Pedro},
\newblock \bibinfo{title}{{Probabilistic Solar Forecasting Using Quantile Regression Models}},
\newblock \bibinfo{journal}{{Energies}} \bibinfo{volume}{10} (\bibinfo{year}{2017}). \DOIprefix\doi{10.3390/en10101591}.
\bibitem[{Oreshkin et~al.(2019)Oreshkin, Carpov, Chapados, and Bengio}]{nbeats}
\bibinfo{author}{B.~N. Oreshkin}, \bibinfo{author}{D.~Carpov}, \bibinfo{author}{N.~Chapados}, \bibinfo{author}{Y.~Bengio},
\newblock \bibinfo{title}{{N-BEATS:} neural basis expansion analysis for interpretable time series forecasting},
\newblock \bibinfo{journal}{CoRR} \bibinfo{volume}{abs/1905.10437} (\bibinfo{year}{2019}). \href{http://arxiv.org/abs/1905.10437}{{\tt arXiv:1905.10437}}.
\bibitem[{Cannon(2011)}]{Cannon2011}
\bibinfo{author}{A.~J. Cannon},
\newblock \bibinfo{title}{Quantile regression neural networks: Implementation in r and application to precipitation downscaling},
\newblock \bibinfo{journal}{Computers \& Geosciences} \bibinfo{volume}{37} (\bibinfo{year}{2011}) \bibinfo{pages}{1277--1284}. \DOIprefix\doi{10.1016/j.cageo.2010.07.005}.
\bibitem[{Bremnes(2020)}]{Bremnes2020}
\bibinfo{author}{J.~B. Bremnes},
\newblock \bibinfo{title}{Ensemble postprocessing using quantile function regression based on neural networks and bernstein polynomials},
\newblock \bibinfo{journal}{Monthly Weather Review} \bibinfo{volume}{148} (\bibinfo{year}{2020}) \bibinfo{pages}{403 -- 414}. \DOIprefix\doi{10.1175/MWR-D-19-0227.1}.
\bibitem[{{Energidataservice}(2024)}]{energidataservice_website}
\bibinfo{author}{{Energidataservice}}, \bibinfo{title}{Energidataservice: The danish energy data service}, \bibinfo{howpublished}{\url{https://www.energidataservice.dk}}, \bibinfo{year}{2024}. \bibinfo{note}{Accessed: 2024-07-01}.
\bibitem[{Andresen et~al.(2015)Andresen, Søndergaard, and Greiner}]{ANDRESEN2015}
\bibinfo{author}{G.~B. Andresen}, \bibinfo{author}{A.~A. Søndergaard}, \bibinfo{author}{M.~Greiner},
\newblock \bibinfo{title}{Validation of danish wind time series from a new global renewable energy atlas for energy system analysis},
\newblock \bibinfo{journal}{Energy} \bibinfo{volume}{93} (\bibinfo{year}{2015}) \bibinfo{pages}{1074--1088}.
\bibitem[{Energistyrelsen(2023)}]{windenergistyrelsen}
\bibinfo{author}{Energistyrelsen}, \bibinfo{title}{Data: Oversigt over energisektoren}, \bibinfo{year}{2023}. \URLprefix \url{https://ens.dk/service/statistik-data-noegletal-og-kort/data-oversigt-over-energisektoren}, \bibinfo{note}{accessed: 2024-11-28}.
\bibitem[{ECMWF(2023)}]{ens}
\bibinfo{author}{ECMWF}, \bibinfo{title}{Ens - ensemble forecasts}, \bibinfo{year}{2023}. \URLprefix \url{https://confluence.ecmwf.int/display/FUG/Section+2.1.2.1+Medium+Range+Ensemble+forecasts}, \bibinfo{note}{accessed: 2024-07-05}.
\bibitem[{{Energinet}(2024)}]{energinet_newdeal}
\bibinfo{author}{{Energinet}}, \bibinfo{title}{Ny model for modhandel sikrer ikke blot mere grøn energi men også store økonomiske gevinster}, \bibinfo{year}{2024}. \URLprefix \url{https://energinet.dk/om-nyheder/nyheder/2024/07/11/ny-model-for-modhandel-sikrer-ikke-blot-mere-gron-energi-men-ogsa-store-okonomiske-gevinster/}, \bibinfo{note}{accessed: 2024-10-29}.
\bibitem[{Nielsen(1999)}]{Nielsen1999}
\bibinfo{author}{H.~Nielsen}, \bibinfo{title}{Algorithms for linear optimization, an introduction}, \bibinfo{year}{1999}. \URLprefix \url{http://www2.imm.dtu.dk/pubdb/views/publication_details.php?id=654}, \bibinfo{note}{course note for the DTU course: Optimization and Data Fitting, vol. 2}.
\bibitem[{M{\o}ller et~al.(2006)M{\o}ller, Nielsen, and Madsen}]{taqrreport}
\bibinfo{author}{J.~M{\o}ller}, \bibinfo{author}{H.~Nielsen}, \bibinfo{author}{H.~Madsen}, \bibinfo{title}{Algorithms for Adaptive Quantile Regression - and a Matlab Implementation}, \bibinfo{year}{2006}.
\bibitem[{Finxter(2023)}]{finxter2023transformerlstm}
\bibinfo{author}{Finxter}, \bibinfo{title}{Transformer vs lstm: A helpful illustrated guide}, \bibinfo{howpublished}{\url{https://blog.finxter.com/transformer-vs-lstm-a-helpful-illustrated-guide/}}, \bibinfo{year}{2023}. \bibinfo{note}{Accessed: 2024-10-30}.
\bibitem[{by~Typeset(2023)}]{scispace2023cases}
\bibinfo{author}{S.~by~Typeset}, \bibinfo{title}{When lstms are better than transformers?}, \bibinfo{howpublished}{\url{https://typeset.io/papers/when-lstms-are-better-than-transformers}}, \bibinfo{year}{2023}. \bibinfo{note}{Accessed: 2024-10-30}.
\bibitem[{Hochreiter and Schmidhuber(1997)}]{lstm}
\bibinfo{author}{S.~Hochreiter}, \bibinfo{author}{J.~Schmidhuber},
\newblock \bibinfo{title}{Long short-term memory},
\newblock \bibinfo{journal}{Neural computation} \bibinfo{volume}{9} (\bibinfo{year}{1997}) \bibinfo{pages}{1735--80}. \DOIprefix\doi{10.1162/neco.1997.9.8.1735}.
\bibitem[{Gers et~al.(1999)Gers, Schmidhuber, and Cummins}]{forgetgate}
\bibinfo{author}{F.~Gers}, \bibinfo{author}{J.~Schmidhuber}, \bibinfo{author}{F.~Cummins},
\newblock \bibinfo{title}{Learning to forget: continual prediction with lstm},
\newblock in: \bibinfo{booktitle}{1999 Ninth International Conference on Artificial Neural Networks ICANN 99. (Conf. Publ. No. 470)}, volume~\bibinfo{volume}{2}, \bibinfo{year}{1999}, pp. \bibinfo{pages}{850--855 vol.2}. \DOIprefix\doi{10.1049/cp:19991218}.
\bibitem[{Olah(2015)}]{olah2015understanding}
\bibinfo{author}{C.~Olah}, \bibinfo{title}{Understanding lstm networks}, \bibinfo{year}{2015}. \URLprefix \url{http://colah.github.io/posts/2015-08-Understanding-LSTMs/}, \bibinfo{note}{accessed: 2024-11-08}.
\bibitem[{Abadi and et~al.(2015)}]{tensorflow}
\bibinfo{author}{M.~Abadi}, \bibinfo{author}{et~al.}, \bibinfo{title}{Tensorflow: Large-scale machine learning on heterogeneous distributed systems}, \bibinfo{howpublished}{\url{https://www.tensorflow.org/}}, \bibinfo{year}{2015}. \bibinfo{note}{Accessed: 2024-06-13}.
\bibitem[{Mozer(1995)}]{bppt1}
\bibinfo{author}{M.~Mozer},
\newblock \bibinfo{title}{A focused backpropagation algorithm for temporal pattern recognition},
\newblock \bibinfo{journal}{Complex Systems} \bibinfo{volume}{3} (\bibinfo{year}{1995}).
\bibitem[{Robinson and Fallside(1987)}]{bppt2}
\bibinfo{author}{A.~J. Robinson}, \bibinfo{author}{F.~Fallside}, \bibinfo{title}{The Utility Driven Dynamic Error Propagation Network}, \bibinfo{type}{Technical Report} \bibinfo{number}{CUED/F-INFENG/TR.1}, Engineering Department, Cambridge University, \bibinfo{address}{Cambridge, UK}, \bibinfo{year}{1987}.
\bibitem[{Werbos(1988)}]{bppt3}
\bibinfo{author}{P.~J. Werbos},
\newblock \bibinfo{title}{Generalization of backpropagation with application to a recurrent gas market model},
\newblock \bibinfo{journal}{Neural Networks} \bibinfo{volume}{1} (\bibinfo{year}{1988}) \bibinfo{pages}{339--356}. \DOIprefix\doi{https://doi.org/10.1016/0893-6080(88)90007-X}.
\bibitem[{Paszke and et~al.(2019)}]{pytorch}
\bibinfo{author}{A.~Paszke}, \bibinfo{author}{et~al.}, \bibinfo{title}{Pytorch: An imperative style, high-performance deep learning library}, \bibinfo{year}{2019}. \URLprefix \url{http://papers.neurips.cc/paper/9015-pytorch-an-imperative-style-high-performance-deep-learning-library.pdf}.
\bibitem[{Koenker and {Bassett Jr.}(1978)}]{Koenker1978}
\bibinfo{author}{R.~Koenker}, \bibinfo{author}{G.~{Bassett Jr.}},
\newblock \bibinfo{title}{Regression quantiles},
\newblock \bibinfo{journal}{Econometrica} \bibinfo{volume}{46} (\bibinfo{year}{1978}) \bibinfo{pages}{33--50}. \DOIprefix\doi{10.2307/1913643}.
\bibitem[{Koenker(2005)}]{Koenker2005}
\bibinfo{author}{R.~Koenker}, \bibinfo{title}{Quantile Regression}, Econometric Society Monographs, \bibinfo{publisher}{Cambridge University Press}, \bibinfo{year}{2005}.
\bibitem[{Van~de Geer(2003)}]{adaptiveqr1}
\bibinfo{author}{S.~A. Van~de Geer},
\newblock \bibinfo{title}{Adaptive quantile regression},
\newblock in: \bibinfo{editor}{M.~G. Akritas}, \bibinfo{editor}{D.~N. Politis} (Eds.), \bibinfo{booktitle}{Recent Advances and Trends in Nonparametric Statistics}, \bibinfo{publisher}{Elsevier Science B.V.}, \bibinfo{year}{2003}, pp. \bibinfo{pages}{235--248}.
\bibitem[{Alhamzawi et~al.(2012)Alhamzawi, Yu, and Benoit}]{adaptiveqr2}
\bibinfo{author}{R.~Alhamzawi}, \bibinfo{author}{K.~Yu}, \bibinfo{author}{D.~Benoit},
\newblock \bibinfo{title}{Bayesian adaptive lasso quantile regression},
\newblock \bibinfo{journal}{Statistical Modelling} \bibinfo{volume}{12} (\bibinfo{year}{2012}). \DOIprefix\doi{10.1177/1471082X1101200304}.
\bibitem[{Chen(2004)}]{adaptiveqr3}
\bibinfo{author}{C.~Chen}, \bibinfo{title}{Theory and Applications of Recent Robust Methods}, Statistics for Industry and Technology, \bibinfo{edition}{1} ed., \bibinfo{publisher}{Birkhäuser Basel}, \bibinfo{address}{Basel, Switzerland}, \bibinfo{year}{2004}. \DOIprefix\doi{10.1007/978-3-0348-7958-3}.
\bibitem[{Dantzig et~al.(1954)Dantzig, Orden, and Wolfe}]{dantzig1954simplex}
\bibinfo{author}{G.~B. Dantzig}, \bibinfo{author}{A.~Orden}, \bibinfo{author}{P.~Wolfe},
\newblock \bibinfo{title}{The generalized simplex method for minimizing a linear form under linear inequality restraints},
\newblock \bibinfo{journal}{Technical Report, ASTIA}  (\bibinfo{year}{1954}). \bibinfo{note}{Rev. 5 April 1954}.
\bibitem[{Meinshausen(2006)}]{qrfMein2006}
\bibinfo{author}{N.~Meinshausen},
\newblock \bibinfo{title}{Quantile regression forests},
\newblock \bibinfo{journal}{Journal of Machine Learning Research} \bibinfo{volume}{7} (\bibinfo{year}{2006}) \bibinfo{pages}{983--999}.
\bibitem[{Friedman(1999)}]{friedman1999stochastic}
\bibinfo{author}{J.~H. Friedman},
\newblock \bibinfo{title}{Stochastic gradient boosting},
\newblock \bibinfo{journal}{Stanford University Technical Report}  (\bibinfo{year}{1999}).
\bibitem[{Friedman(2001)}]{friedman2001greedy}
\bibinfo{author}{J.~H. Friedman},
\newblock \bibinfo{title}{Greedy function approximation: A gradient boosting machine},
\newblock \bibinfo{journal}{Annals of Statistics} \bibinfo{volume}{29} (\bibinfo{year}{2001}) \bibinfo{pages}{1189--1232}.
\bibitem[{Hoyer(2015)}]{properscoring}
\bibinfo{author}{S.~Hoyer}, \bibinfo{title}{Properscoring: Proper scoring rules in python}, \bibinfo{year}{2015}. \URLprefix \url{https://pypi.org/project/properscoring/}.
\bibitem[{Bjerreg{\aa}rd et~al.(2021)Bjerreg{\aa}rd, M{\o}ller, and Madsen}]{mathiasAI}
\bibinfo{author}{M.~B. Bjerreg{\aa}rd}, \bibinfo{author}{J.~K. M{\o}ller}, \bibinfo{author}{H.~Madsen},
\newblock \bibinfo{title}{An introduction to multivariate probabilistic forecast evaluation},
\newblock \bibinfo{journal}{Energy and AI} \bibinfo{volume}{4} (\bibinfo{year}{2021}) \bibinfo{pages}{100058}. \DOIprefix\doi{https://doi.org/10.1016/j.egyai.2021.100058}.
\bibitem[{Pedregosa et~al.(2011)Pedregosa, Varoquaux, Gramfort, Michel, Thirion, Grisel, Blondel, Prettenhofer, Weiss, Dubourg, Vanderplas, Passos, Cournapeau, Brucher, Perrot, and Duchesnay}]{scikitlearn}
\bibinfo{author}{F.~Pedregosa}, \bibinfo{author}{G.~Varoquaux}, \bibinfo{author}{A.~Gramfort}, \bibinfo{author}{V.~Michel}, \bibinfo{author}{B.~Thirion}, \bibinfo{author}{O.~Grisel}, \bibinfo{author}{M.~Blondel}, \bibinfo{author}{P.~Prettenhofer}, \bibinfo{author}{R.~Weiss}, \bibinfo{author}{V.~Dubourg}, \bibinfo{author}{J.~Vanderplas}, \bibinfo{author}{A.~Passos}, \bibinfo{author}{D.~Cournapeau}, \bibinfo{author}{M.~Brucher}, \bibinfo{author}{M.~Perrot}, \bibinfo{author}{E.~Duchesnay},
\newblock \bibinfo{title}{Scikit-learn: Machine learning in {P}ython},
\newblock \bibinfo{journal}{Journal of Machine Learning Research} \bibinfo{volume}{12} (\bibinfo{year}{2011}) \bibinfo{pages}{2825--2830}.
\bibitem[{Jørgensen(2024)}]{bastianmsc}
\bibinfo{author}{B.~S. Jørgensen}, \bibinfo{title}{Real-Time Forecasting in Renewable Energy Production: Improving Wind and Solar Forecasts through Advanced Error Correction Methods}, Master's thesis, DTU Department of Applied Mathematics and Computer Science, \bibinfo{year}{2024}. \URLprefix \url{http://people.compute.dtu.dk/bassc/}.
\bibitem[{M{\o}ller et~al.(2016)M{\o}ller, Zugno, and Madsen}]{sde1}
\bibinfo{author}{J.~K. M{\o}ller}, \bibinfo{author}{M.~Zugno}, \bibinfo{author}{H.~Madsen},
\newblock \bibinfo{title}{Probabilistic forecasts of wind power generation by stochastic differential equation models},
\newblock \bibinfo{journal}{Journal of Forecasting} \bibinfo{volume}{35} (\bibinfo{year}{2016}) \bibinfo{pages}{189--205}. \DOIprefix\doi{https://doi.org/10.1002/for.2367}.
\bibitem[{Li et~al.(2024)Li, Liu, Tian, Wang, Wang, Jin, Wu, Tan, Lin, Liu, Sun, and Li}]{li2024switchemafreelunch}
\bibinfo{author}{S.~Li}, \bibinfo{author}{Z.~Liu}, \bibinfo{author}{J.~Tian}, \bibinfo{author}{G.~Wang}, \bibinfo{author}{Z.~Wang}, \bibinfo{author}{W.~Jin}, \bibinfo{author}{D.~Wu}, \bibinfo{author}{C.~Tan}, \bibinfo{author}{T.~Lin}, \bibinfo{author}{Y.~Liu}, \bibinfo{author}{B.~Sun}, \bibinfo{author}{S.~Z. Li}, \bibinfo{title}{Switch ema: A free lunch for better flatness and sharpness}, \bibinfo{year}{2024}. \href{http://arxiv.org/abs/2402.09240}{{\tt arXiv:2402.09240}}.
\bibitem[{Finnah(2022)}]{finnah2022optimal}
\bibinfo{author}{B.~Finnah},
\newblock \bibinfo{title}{Optimal bidding functions for renewable energies in sequential electricity markets},
\newblock \bibinfo{journal}{OR Spectrum} \bibinfo{volume}{44} (\bibinfo{year}{2022}) \bibinfo{pages}{1--27}. \DOIprefix\doi{10.1007/s00291-021-00646-9}.
\bibitem[{Arat(2019)}]{arat2019dimensions}
\bibinfo{author}{M.~M. Arat}, \bibinfo{title}{Dimensions of matrices in an lstm cell}, \bibinfo{year}{2019}. \URLprefix \url{https://mmuratarat.github.io/2019-01-19/dimensions-of-lstm}, \bibinfo{note}{accessed: 2024-11-08}.

\end{thebibliography}

\appendix
\onecolumn
\section{Reliability plot} \label{sec:appendix_rel}
\begin{figure*}[!h]
    \centering
    \includegraphics[width=1\linewidth]{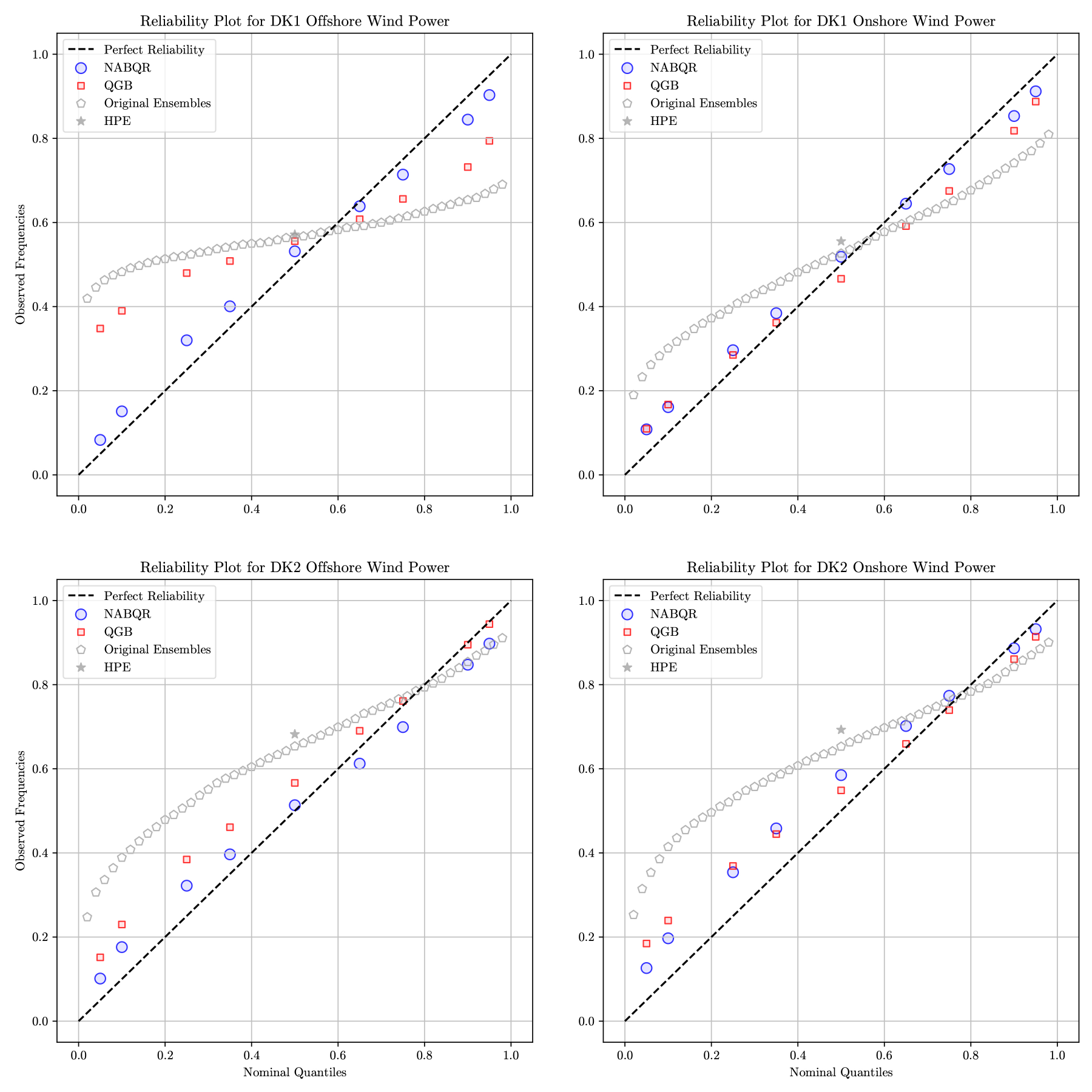}
    \caption{Reliability plot for NABQR, QGB and the original ensembles for all four areas.}
    \label{fig:relplot}
\end{figure*}

\onecolumn
\section{Dimensionality overview for LSTM network in TensorFlow}
Let $B$ be the batch size, $F$ the number of features and $U$ the number of units in an LSTM cell, then we have the following dimensions, which are useful in relation to the equations governing the LSTM network in this paper.
\begin{table}[!ht]
\centering
\caption{Dimension in LSTM network. $B$ is batch size, $F$ number of features and $U$ is number of units.}
\begin{tabular}{|c|c|}
\hline
\textbf{Variable} & \textbf{Dimensions} \\
\hline
$x_t$ & $\mathbb{R}^{B \times F}$ \\
$h_{t-1}$ & $\mathbb{R}^{B \times U}$ \\
$h_t$ & $\mathbb{R}^{B \times U}$ \\
$C_{t-1}, C_t, \tilde{C}_t$ & $\mathbb{R}^{B \times U}$ \\
\hline
$W_i$ & $\mathbb{R}^{(F+U) \times U}$ \\
$W_c$ & $\mathbb{R}^{(F+U) \times U}$ \\
$W_f$ & $\mathbb{R}^{(F+U) \times U}$ \\
$W_o$ & $\mathbb{R}^{(F+U) \times U}$ \\
\hline
$b_i$ & $\mathbb{R}^{U}$ \\
$b_c$ & $\mathbb{R}^{U}$ \\
$b_f$ & $\mathbb{R}^{U}$ \\
$b_o$ & $\mathbb{R}^{U}$ \\
\hline
$i_t$ & $\mathbb{R}^{B \times U}$ \\
$f_t$ & $\mathbb{R}^{B \times U}$ \\
$h_t$ & $\mathbb{R}^{B \times U}$ \\
$o_t$ & $\mathbb{R}^{B \times U}$ \\
\hline
\end{tabular}
\label{tab:dimensionLSTM}
\end{table}

However, the calculations behind the scenes are best understood considering the following:
\[
\begin{aligned}
i_t &= \sigma\bigl(W_{xi} X_t + W_{hi} h_{t-1} + b_i\bigr) \\
f_t &= \sigma\bigl(W_{xf} X_t + W_{hf} h_{t-1} + b_f\bigr) \\
\tilde{C}_t &= \tanh\bigl(W_{xc} X_t + W_{hc} h_{t-1} + b_c\bigr) \\
C_t &= f_t \circ C_{t-1} \;+\; i_t \circ \tilde{C}_t \\
o_t &= \sigma\bigl(W_{xo} X_t + W_{ho}^\top h_{t-1} + b_o\bigr) \\
h_t &= o_t \circ \tanh(C_t)
\end{aligned}
\]

\noindent\textit{Note:} The bias vectors, $b_i, b_f, b_c, b_o$, is broadcasted so that dimensions match. 

The weight matrices \(W_{xi}, W_{xf}, W_{xc}, W_{xo}\) correspond to the input vector \(X_t\) for each gate and the candidate cell input. Similarly, \(W_{hi}, W_{hf}, W_{hc}, W_{ho}\) are the corresponding weight matrices for the previous hidden state \(h_{t-1}\). 

For further information on the implementation of these large-scale matrix multiplications and related computations in TensorFlow, see \citep{arat2019dimensions}.


\end{document}